*Research Article*

# The Role of Visual Features in Text-Based CAPTCHAs: An fNIRS Study for Usable Security

**Emre Mülazimoğlu** ,[1] **Murat P. Çakır** ,[2] **and Cengiz Acartürk** [1,2]

[1]*Cyber Security Department, Middle East Technical University, Ankara 06800, Turkey*
[2]*Department of Cognitive Science, Middle East Technical University, Ankara 06800, Turkey*

Correspondence should be addressed to Cengiz Acartürk; acarturk@metu.edu.tr

To mitigate dictionary attacks or similar undesirable automated attacks to information systems, developers mostly prefer using CAPTCHA challenges as Human Interactive Proofs (HIPs) to distinguish between human users and scripts. Appropriate use of CAPTCHA requires a setup that balances between robustness and usability during the design of a challenge. The previous research reveals that most usability studies have used accuracy and response time as measurement criteria for quantitative analysis. The present study aims at applying optical neuroimaging techniques for the analysis of CAPTCHA design. The functional Near-Infrared Spectroscopy technique was used to explore the hemodynamic responses in the prefrontal cortex elicited by CAPTCHA stimulus of varying types. The findings suggest that regions in the left and right dorsolateral and right dorsomedial prefrontal cortex respond to the degrees of line occlusion, rotation, and wave distortions present in a CAPTCHA. The systematic addition of the visual effects introduced nonlinear effects on the behavioral and prefrontal oxygenation measures, indicative of the emergence of Gestalt effects that might have influenced the perception of the overall CAPTCHA figure.

## 1. Introduction

CAPTCHA challenges have a vital role as a countermeasure of automated web application attacks. According to OWASP's (Open Web Application Security Project) list of top 10 web application security risks, brute force and dictionary attacks are reported as the second most critical attack by cybersecurity experts in 2017 [1]. To mitigate dictionary attacks, more specifically script-automated dictionary attacks, software developers usually prefer using CAPTCHA challenges. CAPTCHA challenges comprise a specific family of Human Interactive Proofs (HIPs), which aim at distinguishing genuine human users from automated scripts. CAPTCHA is an acronym for "Completely Automated Public Turing Test To Tell Computers and Humans Apart." The term was first coined two decades ago [2]. A CAPTCHA enforces users to solve a given challenge to prove that they are, indeed, human. The challenges come in a variety of forms. For instance, in text-based form, the challenge is usually to enter a set of alphanumeric characters from a keyboard, displayed in a distorted image on display. Similarly, in image-based form, the challenge may consist of choosing images that contain a specific object from a set of images.

Chew and Tygar [3] outlined three properties that must be satisfied by a CAPTCHA challenge: "1) easy for humans to solve, (2) hard for scripts to solve, and (3) easy for tester software to generate and grade" (p. 268). The first two properties address a balance between usability and robustness in design. Accordingly, a CAPTCHA must be usable by humans, and at the same time, it must be robust against automated scripts. To assess the robustness, researchers have developed various attack methodologies and tested them against available challenge designs. The methodologies have been subject to interdisciplinary development on various fronts, including image processing and computer vision and pattern recognition, and machine learning [4].

As in other domains subject to intrusion by artificial intelligence, CAPTCHA is a challenge between human users and computers and a continuous game of cops and robbers between software developers and criminal hackers that perform dictionary attacks against authentication systems. So, for the last two decades, once a new challenge design became solvable by programming or compromised by any automated means, researchers developed more robust security features within the same type or novel type of the challenge [4]. In this article, we aim to contribute to the study of the CAPTCHA concept by presenting a report of an experimental study that investigated users' brain activity during their CAPTCHA solving processes.

The findings of the present study have an impact on usability since difficult challenges make users less satisfied due to reduced usability, i.e., a higher probability of failure or low familiarity. From the software developer's perspective, performance results, such as response time and accuracy, give valuable information about user experience during authentication utilizing CAPTCHA interfaces. However, the recent advances in neuroscience allow us to go beyond performance measures [5]. Within the context of end-user authentication, neuroimaging is a promising method for developing more direct cognitive workload measures on the user side than behavioral performance. In the present study, we employ optical neuroimaging to study cognitive workload during CAPTCHA solving tasks.

Optical neuroimaging techniques have been used for the study of cognitive processes under various contextual environments, including driving [6], subject classification in memory tasks [7], comparison of brain activity in motor imagery and motor movement [8], and usability and mental workload analysis in order to understand hemodynamic responses in the brain [9]. Recently, optical neuroimaging, and more generally neuroimaging, has limited applications in cybersecurity research. In particular, to the best of our knowledge, no research has been conducted that focuses on using neuroimaging techniques during CAPTCHA solving.

In the present study, we used Functional Near-Infrared Spectroscopy (fNIRS) to monitor brain responses at the prefrontal cortex during CAPTCHA solving. fNIRS is a neuroimaging modality that enables continuous, noninvasive, and portable monitoring of blood oxygenation changes and blood volume related to human brain function [10]. Neuronal activity is determined concerning oxygenation changes since variation in cerebral hemodynamics is related to functional brain activity through a mechanism known as neurovascular coupling [11]. Despite its limitations in spatial resolution and the depth of the cortical tissue that can be monitored, the portability, ease of use, and higher temporal resolution make fNIRS advantageous compared to other modalities such as fMRI and PET, and fNIRS can also monitor hemodynamic responses. In contrast to electrophysiological modalities such as EEG/MEG, fNIRS cannot provide comparable temporal resolution due to the slow accumulation of hemodynamic response as a consequence of neural activity. However, fNIRS provides better spatial resolution, especially in contrast to EEG, since the propagation of optical signals is less influenced by the brain tissue than local field potentials. In short, fNIRS provides a good balance in terms of spatial/temporal resolution, portability, and cost for conducting applied cognitive neuroscience studies in human-computer interaction and cybersecurity settings. Therefore, this study aims to explore the plausibility of using a portable fNIRS system for monitoring brain responses elicited by a cybersecurity application, namely, CAPTCHA solving.

*1.1. CAPTCHA Solving as a Cognitive Process.* Various visual features are employed for designing text-based CAPTCHAs that present distorted alphanumerical characters, such as color, font, character type, length, and text content (e.g., word or nonword) [12, 13]. The development of the visual features of CAPTCHA design has been accomplished by the development of the methodologies that perform automated solutions.

A review of the previous studies reveals that most of the usability studies on CAPTCHAs have used accuracy and response time (of solving) as measurement criteria for a quantitative evaluation of CAPTCHA design's effectiveness. For instance, Bursztein et al. [14] developed a single-step approach that used machine learning algorithms to solve a CAPTCHA automatically. Since then, application developers have started to use more challenging CAPTCHA types to bypass automated systems, such as image-based CAPTCHAs, in addition to text-based ones. However, due to their deployment complexity for implementation and the dependency of third-party cloud services for API (Application Programming Interface) implementation, text-based CAPTCHAs are still widely used today. As a mitigation method against cybersecurity attacks, CAPTCHAs provide an additional defense-in-depth mechanism besides traditional challenges that accompany other countermeasures, such as rate-limiting or interaction detection that provides secure authentication in information-sensitive applications.

A further investigation of the visual features employed in CAPTCHA design reveals a set of studies investigating CAPTCHA design from the perspective of usability. For instance, Chellapilla et al. [15] analyzed a set of visual features, including scaling, rotation, global and local warping, and arcs, using accuracy as a usability metric. Bursztein et al. [16] investigated 13 different popular text-based and eight different audio-based types of challenges, including the CAPTCHAs used by Baidu, Google, eBay, and Slash. They employed the solution time and accuracy as the usability metrics. Others employed the number of attempts to solve the challenge, in addition to response time and accuracy, to analyze the role of length, size, language and distortion level [13], letter case, and orthographic properties of lowercase letters on usability [17, 18].

The investigation of behavioral performance (response time and accuracy) in CAPTCHA solving has been subject to large-scale studies, in addition to small-scale laboratory experiments. For instance, an experiment was conducted on Amazon's Mechanical Turk with over 27,000 respondents [14]. They asked the respondents to solve nearly a total of a million CAPTCHAs. The main purpose of the study was to redesign Google's reCaptcha v1. The focus was to analyze the interaction of a set of visual features and analyze antisegmentation and antirecognition features in isolation (approximately 20 features including content types), which eventually led to large stimuli set. In particular, they mainly used 6–8 digits, overlapping (with a line for backup), up to 20-degree random rotation, length and font size randomization, and sinusoidal waving with a specific configuration, as the visual features.

In the present study, we focused on investigating three major visual features (line, wave, and rotation) and their interaction, which were also used by [14]. To limit the size of the stimuli, we used 6-digit, numerical CAPTCHAs. In addition to measuring behavioral performance, we employed the fNIRS optical neuroimaging method to measure blood oxygenation changes due to neuronal activity in the prefrontal cortex while participants were solving CAPTCHAs of various difficulty levels.

Briefly, fNIRS uses specific wavelengths of light in the red and infrared wavelengths to monitor the relative changes in oxyhemoglobin (HbO) and deoxyhemoglobin (HbR) concentrations in the capillary beds within cortical tissue. Neuronal activity is determined with respect to changes in oxygenation since variation in cerebral hemodynamics is related to functional brain activity through a mechanism known as neurovascular coupling [19, 20]. Neural activity is an energy-intensive process that requires oxygen supplied by the vascular system to facilitate glucose metabolism. When a neural population becomes active, it first consumes the oxygen available in the vicinity. It then triggers a rush of oxygenated blood towards that region; a phenomenon called the hemodynamic response [11]. fNIRS systems typically feature LED or laser-based light sources to illuminate the tissue with photons of specific wavelengths and detectors positioned at specific distances from the light sources to monitor the intensity of returning light. Photons that travel through tissue are mainly subjected to absorption and scattering [21]. Within the 700–900 nm range, oxy- and deoxyhemoglobin molecules are the strongest absorbers, whereas the skin, tissue, and bone structures are mainly transparent. Moreover, the absorption spectra of oxy- and deoxyhemoglobin are sufficiently different from each other in this optical window, making it possible to estimate their relative concentrations as a function of light intensity changes across two wavelengths by using the modified Beer-Lambert law [22].

Several fNIRS and fMRI studies have investigated cortical hemodynamics changes during tasks that can be associated with CAPTCHA solving, such as letter recognition, letter/number copying, mental rotation, and anagram solving tasks. In an fMRI study on the orthographic encoding of letters, symbols, and digits, Carrieras et al. [23] employed a string comparison task where participants decided whether two four-character long strings displayed in succession were the same or different, where the different cases required a further distinction based on whether a pair of characters were transposed or replaced in the successive string. They reported increased activity in the left inferior frontal gyrus, particularly for processing letters as opposed to digits/symbols, especially when the stimuli were harder to process (i.e., stimuli that elicited higher error rate or longer response time), suggesting the possible involvement of the left-lateralized language network for letter processing and the top-down attentional mechanism for the resolution of more complex cases. Similarly, an fNIRS study on a letter copying task, including retyping of a given number of letters or numbers on the screen, reported that both types of stimuli recruited increased activity across the bilateral middle and inferior frontal gyri along with regions at the superior parietal lobule and superior temporal gyrus [24]. Mental rotation is another related task where participants are given two shapes and asked to decide if the second image could be obtained by rotating the first one or not. Several variants of the task have been investigated in the neuroimaging literature, including 3D blocks, letters, and abstract symbols [25]. The meta-analysis findings suggest that it typically took subjects more time to respond to the task with increasing angular separation between the two images. The task robustly recruits bilateral inferior parietal sulcus and middle and dorsolateral prefrontal cortex with increasing task difficulty [25, 26]. Finally, anagram solving is another task that can be considered relevant. The participants try to find a word using the letters presented to them in mixed order (e.g., atbel to table). An fNIRS study including anagram tasks of varying difficulty reported increased activity in the left dorsolateral and medial prefrontal cortex with further increase in maximum HbO amplitudes in the case of more difficult anagrams [27]. Overall, based on these results, it can be hypothesized that CAPTCHA solving related processes for recognizing letters/digits within distorted images will likely recruit regions in the prefrontal cortex, especially with increasing difficulty. To the best of our knowledge, CAPTCHA solving processes have not been investigated in a neuroimaging study.

Accordingly, the specific purpose of the present study is to investigate the relationship between text-based CAPTCHA solving tasks and the hemodynamic responses in the brain. This investigation aims at evaluating hemodynamic responses as a complementary methodology to the analysis of behavioral variables (accuracy and response time) for the selected visual features. Our further goal is to examine the effects of the selected visual features (line, wave, and rotation) with respect to the prefrontal cortex (PFC) regions. In the present study, the visual features are the independent variables. Response time, accuracy, and changes in oxyhemoglobin and deoxyhemoglobin are the dependent variables. The following section presents the experimental investigation.

## 2. Experiment

*2.1. The Participants.* The experiment was conducted with 25 participants, who were university students. All the participants were right-handed. The mean age of the participants was 25.0 (SD = 2.37, 13 female). Nine participants had eyeglasses, and one of them had contact lens during the experiment. There was no color-blind participant as reported in the demographic data forms. Eight participants were undergraduate students, ten had a Bachelor's degree, and the rest had a Master's degree. All participants were familiar with CAPTCHAs as frequent computer users in their daily lives.

*2.2. Material Design and Stimuli.* Numerous visual features have been used in text-based CAPTCHA design, such as the addition of visual clutter as noise, collapsing alphanumerical characters, and skewing the characters around an axis. Following up the literature review, we selected three visual features as design parameters for developing the stimuli. In particular, we selected line, wave, and rotation, which were also the main visual features of the final design of Google's reCaptcha v1 [14]. We used the three visual features for generating CAPTCHAs with three levels of difficulty. For all three visual features, the easiest level meant no application of the features onto the CAPTCHA. The second level of difficulty meant a simple application of the feature onto the CAPTCHA. The third level included the application of the visual feature onto the CAPTCHA vigorously (see Figure 1 for sample stimuli). We expected that the level of difficulty would correlate with the usability negatively.

As for the visual features, we applied the line feature by adding a thin line (Level-2 difficulty) or a thick line (Level-3 difficulty) in the middle of a CAPTCHA passing through all the characters. The second feature was the wave feature with three levels. We relocated the orientation of CAPTCHAs on a sinusoidal line at the second level of difficulty for applying the wave effect. On the third level and the second level, more wave effect was applied on each CAPTCHA character. The last feature was the rotation feature with two levels of difficulty. At the second level of difficulty, we rotated each character independently with the angle randomly selected from an array including {-20, 15, 15, 20} degrees. The rotation feature has no more levels since rotated characters with more degrees start overlapping by dramatically decreasing accuracy. All the three visual features and their difficulty levels are summarized in Table 1.

To analyze the effect of each feature and the difficulty level, all the combinations of these different features were analyzed. A total of $(3 \times 3 \times 2)$ 18 types of text-based CAPTCHA were generated (a complete list of CAPTCHA types is presented in Table 2). The CAPTCHA with no visual feature was chosen as the baseline in the analyses. We used a black foreground, a grey background, and Arial regular true-type font family with 22 font sizes to generate each character of the CAPTCHAs. Six-digit numeric character sets were randomly generated for each participant, so the participants solved CAPTCHAs of random digit combinations. We used an open-source tool, namely, cool-php-captcha (https://github.com/josecl/cool-php-captcha, last retrieved on July 31, 2020.), to generate the stimuli, with code adjustments to match the exact visual features expected in our design.

The findings in the previous studies show that solving a typical text-based CAPTCHA takes 3–7 seconds. A block design was employed where five CAPTCHA challenges in the same type were presented to the participant to detect hemodynamic responses in the prefrontal cortex to each specific combination of CAPTCHA features, as shown in

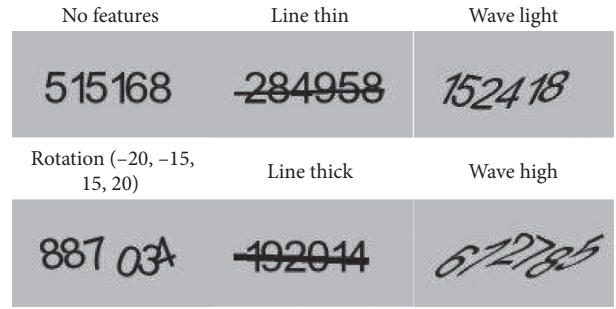

Figure 1: Example images generated for each single CAPTCHA feature.

Figure 2. Between each CAPTCHA in the block, a two-second rest was given. At the beginning of each block, CAPTCHAs without any visual distortions were given as a baseline task. Participants were asked to enter the CAPTCHA text as quickly as possible, but no time-out mechanism was employed. On average, participants completed a block in 47.9 seconds (SD = 12.2).

A total of 18 blocks were designed and displayed to each participant in random order. Between each block, there were 10-second long rest periods. At the beginning of each experiment, a test run block was run to familiarize the participants with the experiment. After the test run and before the main experiment session, a 20-second rest was introduced. So, a total of 19 blocks were displayed during the experiment. The overall block design is presented in Figure 3.

In order to display the stimuli, a web application was developed by the PHP programming language. The application included a user interface that can record all the data needed for response time and accuracy measurements and demographic data of the participants. In addition to typed answers and their correctness, the displayed stimuli were recorded in base64 form for troubleshooting purposes if needed. The web application communicated with the fNIRS device by sending a marker for synchronization. The markers were used to identify the start of a block, the end of baseline CAPTCHAs, and the end of a block.

*2.3. Procedure.* Before the experiment, each participant signed informed consent forms as part of ethical compliance in experimental studies at the host university. Participants filled in a translated version of the Edinburgh Handedness Inventory questionnaire for handedness and a demographic data form [28]. The participants were informed about privacy, the equipment, and the interface. In the experiment session, each participant attempted to solve a total of 133 CAPTCHAs in approximately 20–30 minutes.

Accuracy and response time data were recorded by the native software that displayed the CAPTCHAs to the participants. Hemodynamic activity in each participant's prefrontal cortex was monitored with a continuous wave fNIRS system developed at Drexel University and manufactured and supplied by fNIR Devices LLC (Potomac, MD; http://www.fNIRdevices.com). The system is composed of a

Table 1: Selected features to be analyzed with difficulty levels.

| Features | # of difficulties | Security features defined for each level | | | |
|---|---|---|---|---|---|
| Line | 3 levels | No line | Line thin | Line thick |
| Waving | 3 levels | No wave | Wave light | Wave high |
| Rotation | 2 levels | No rotation | Rotation {−20, −15, 15, 20} | — |

Table 2: Final security features of eighteen types of text-based CAPTCHA used in our experiment.

| Type# | Line features | Waving features | Rotation features |
|---|---|---|---|
| 1 | No line | No wave | No rotation |
| 2 | Line thin | | |
| 3 | Line thick | | |
| 4 | | | Max rotation 20 |
| 5 | | Wave light | |
| 6 | | Wave high | |
| 7 | Line thin | | Max rotation 20 |
| 8 | Line thick | | Max rotation 20 |
| 9 | Line thin | Wave light | |
| 10 | Line thin | Wave high | |
| 11 | Line thick | Wave high | |
| 12 | Line thick | Wave high | |
| 13 | | Wave light | Max rotation 20 |
| 14 | | Wave high | Max rotation 20 |
| 15 | Line thin | Wave light | Max rotation 20 |
| 16 | Line thin | Wave high | Max rotation 20 |
| 17 | Line thick | Wave light | Max rotation 20 |
| 18 | Line thick | Wave high | Max rotation 20 |

Max rotation 20 = rotation (−20, −15, 15, 20).

flexible headpiece, which holds four light sources and ten detectors to obtain oxygenation measures at 16 optodes over the prefrontal cortex; a control box for hardware management; a computer that runs COBI Studio software (Ayaz et al., 2011) for data acquisition (Figure 4). The system records raw light intensity measurements at two wavelengths, namely, 730 nm and 850 nm, and in the ambient mode to detect possible leakages due to poor skin contact. The sensor has a source-detector separation of 2.5 cm, which allows for approximately 1.25 cm penetration depth. This system can monitor changes in relative concentrations of HbO and HbR at a temporal resolution of 2 Hz. The 16 optodes correspond to regions in the Broadmann areas 9, 10, 44, and 45 over the prefrontal cortex [29].

The experiment was conducted in a laboratory setting. The participants used a numpad keyboard for entering numeric CAPTCHA values. The experiment setup is illustrated in Figure 5.

## 3. Results

Twenty-five participants solved 3,325 numeric text-based CAPTCHAs in total during the experiment, including the test run blocks. Test run and baseline CAPTCHA tests were always the two initial stimuli in each block. Accordingly, they were excluded from the analysis reported in this section. Due to a technical problem in one CAPTCHA type shown for one of the participants, the related data were excluded. Finally, we analyzed the remaining 2,160 CAPTCHA results for twenty-four participants. IBM SPSS v25 and fNIRSoft v4.11 [30] were used for the processing and statistical analysis of the collected data.

### 3.1. Behavioral Results (Accuracy and Response Time).
The overall mean accuracy was 78.7% (SD = 28.2%), and the overall mean response time was 5.50 seconds (SD = 2.94). The response time of each solved CAPTCHA was recorded by a JavaScript function called performance.now, which measured the duration between stimuli display (web page loading) and the click action on the submit button. The logs indicated that all participants attempted all presented CAPTCHAs.

Three-way repeated-measures ANOVAs were conducted to compare the effect of the difficulty levels of the visual features (line, wave, and rotation) on accuracy (Figure 6) and response time (Figure 7). The analysis on accuracy levels revealed a significant main effect of line ($F(1.41, 32.4)$ = 225.6, $p < 0.01$, partial $\eta^2 = .91$), wave ($F$ (2, 46) = 39.66, $p < 0.01$, partial $\eta^2 = .63$), and rotation ($F$ (1, 23) = 177.4, $p < 0.01$, partial $\eta^2 = .89$). The interactions of wave and line ($F$ (2.83, 65.00) = 11.82, $p < 0.01$, partial $\eta^2 = .34$), wave and rotation ($F$ (2, 46) = 8.02, $p < 0.01$, partial $\eta^2 = .26$), and line and rotation ($F$ (2, 46) = 32.27, $p < 0.01$, partial $\eta^2 = .58$) were also significant. The three-way interaction was not significant, $F$ (4, 92) = 1.50, $p > 0.05$.

Sidak-corrected post hoc tests found that participants scored significantly lower in the line-3 condition ($M$ = 93.33, $SD$ = 0.94) than in line-2 ($M$ = 89.17, SD = 1.32) and line-1 ($M$ = 53.61, SD = 2.64). Similarly, average accuracy was significantly low for wave-3 ($M$ = 70.56, SD = 1.90) and was significantly high for wave-2 ($M$ = 78.33, SD = 1.69) and wave-1 ($M$ = 87.22, SD = 1.55), respectively.

The ANOVA results on response times revealed a significant main effect of *line* ($F$ (1.13, 26.1) = 87.8, $p < 0.01$, partial $\eta^2 = .79$), wave ($F$ (2, 46) = 29.95, $p < 0.01$, partial $\eta^2 = .57$), and rotation ($F$ (1, 23) = 77.5, $p < 0.01$, partial $\eta^2 = .77$). The interactions of wave and line ($F$ (2.63, 60.59) = 12.22, $p < 0.01$, partial $\eta^2 = .35$), wave and rotation ($F$ (2, 46) = 9.57, $p < 0.01$, partial $\eta^2 = .29$), and line and rotation ($F$ (2, 46) = 26.16, $p < 0.01$, partial $\eta^2 = .53$) were also significant. The three-way interaction was also significant, $F$ (2.52, 57.96) = 3.25, $p < 0.05$, partial $\eta^2 = .12$.

Sidak-corrected post hoc tests found that participants took significantly longer time while responding to line-3 condition ($M$ = 7046.48, SD = 397.27) than that while responding to line-2 ($M$ = 4806.28, SD = 252.64) and line-1 ($M$ = 4697.67, SD = 237.69). Similarly, the average response times observed for wave-3 ($M$ = 5956.96, SD = 354.99) were significantly higher than wave-2 ($M$ = 5776.86, SD = 298.76) and wave-1 ($M$ = 4816.61, SD = 218.93), respectively.

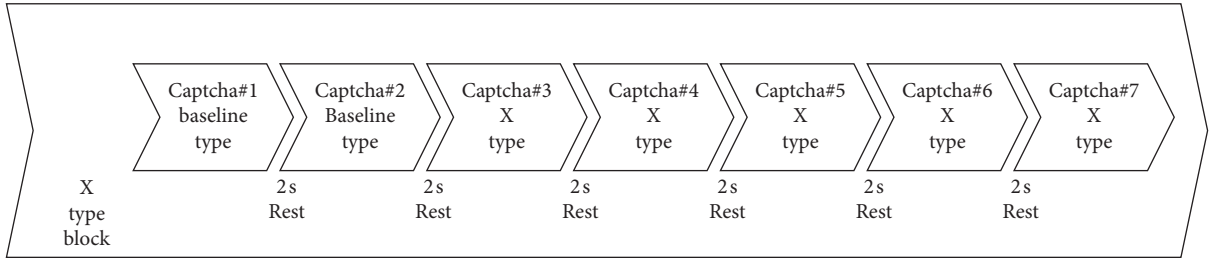

Figure 2: The sequence of CAPTCHA stimuli displayed in one main block design.

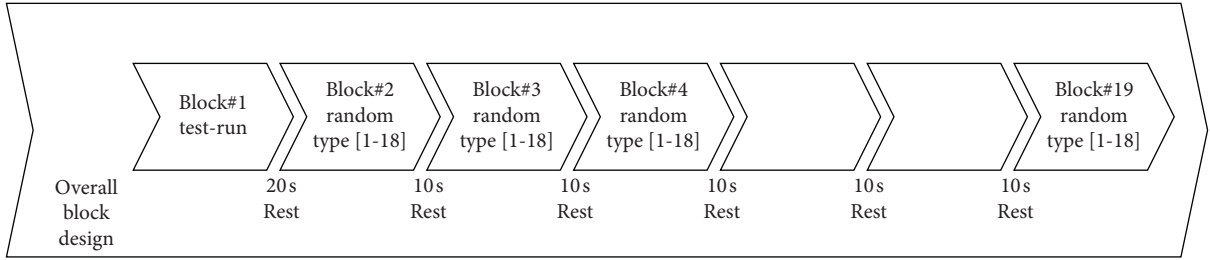

Figure 3: Overall block design used in the experiment.

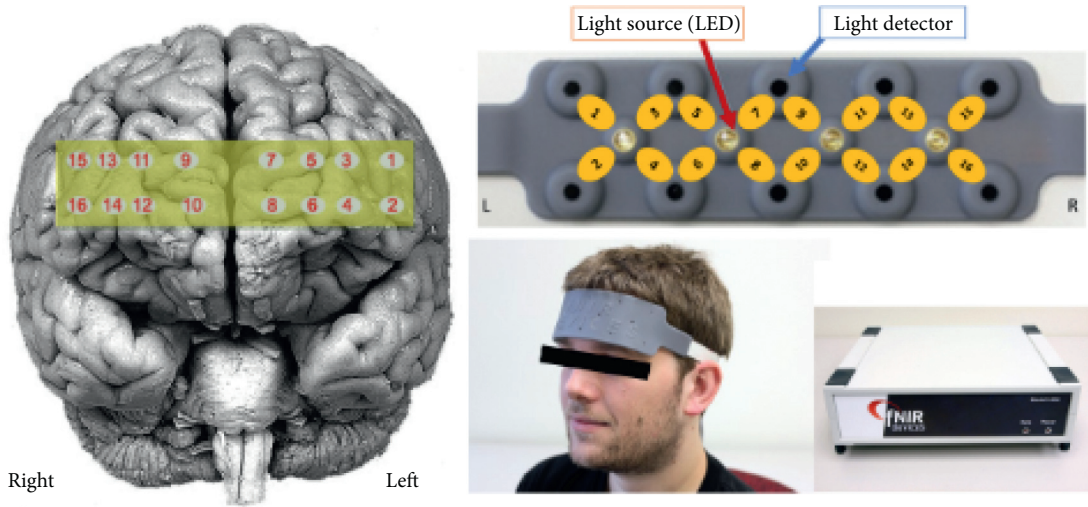

Figure 4: The projection of measurement locations (optodes) on a brain surface image (left), the light detectors and optodes identified on the fNIRS sensor (top-right), the flexible sensor pad placed over the forehead, and the control box (bottom-right).

To sum up, the behavioral results were mainly compatible with the behavioral findings reported in previous research. When no distortions are presented, the accuracy is at the ceiling and the response time average is at the lowest level. When line features are turned on, the most remarkable difference occurs in the case of thick lines. A similar situation applies to the wave distortion, making the most significant impact with its last level when all other features are disabled. The rotation effect also reduces accuracy, increases response time, and amplifies the effects of line occlusion and wave distortion. The decrease in accuracy and the corresponding increase in response time followed a nonlinear trend as participants attempted CAPTCHAs with increasing levels of line, wave, and rotation distortions.

*3.2. Neuroimaging Results (fNIRS).* The preprocessing of collected fNIRS data was performed with the fNIRSoft Professional software [30]. Raw fNIRS data (16 optodes × 2 wavelengths) were low-pass filtered with a finite impulse response, linear phase filter with order 20, and cut-off frequency 0.1 Hz to attenuate the high-frequency noise due to respiration and cardiac cycle effects [31]. Saturated channels (if any) in which light intensity at the detector was higher than the analog-to-digital converter limit were excluded. Such cases typically occur when there is poor skin-detector contact or hair caught between the skin and the detector. Artifacts because of motion were detected and excluded by applying the sliding windows motion artifact filter [32].

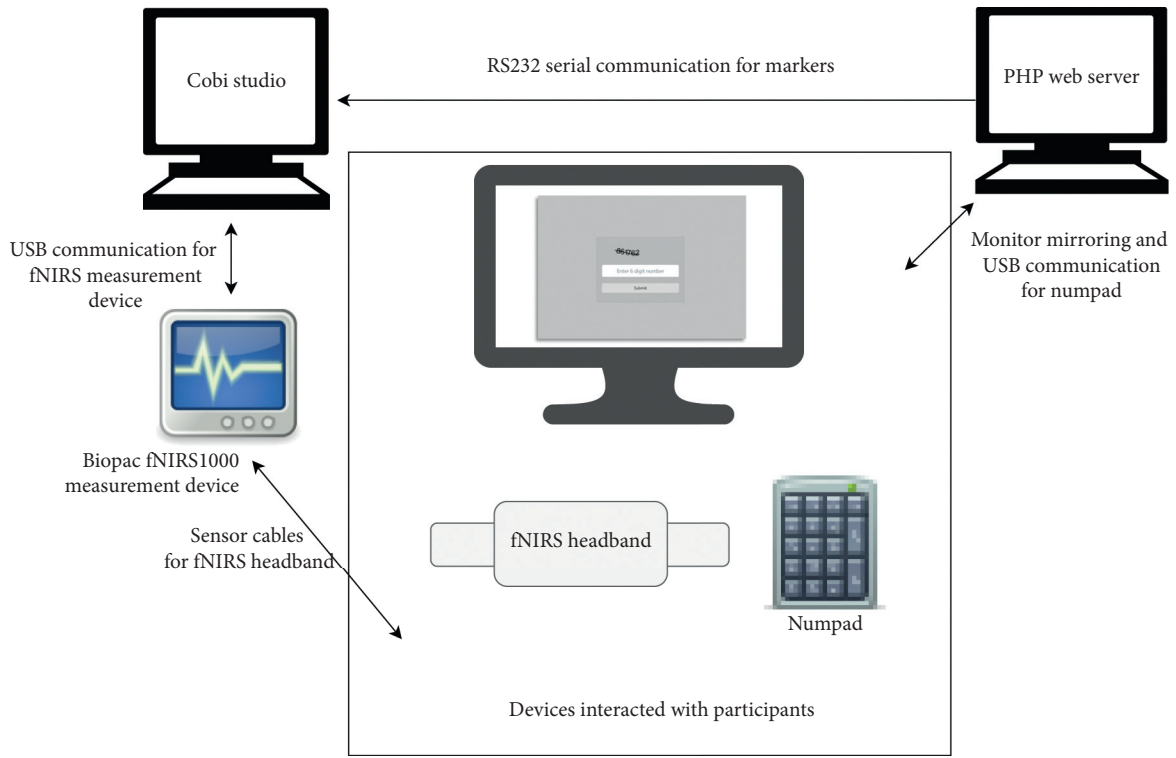

Figure 5: Experiment setup.

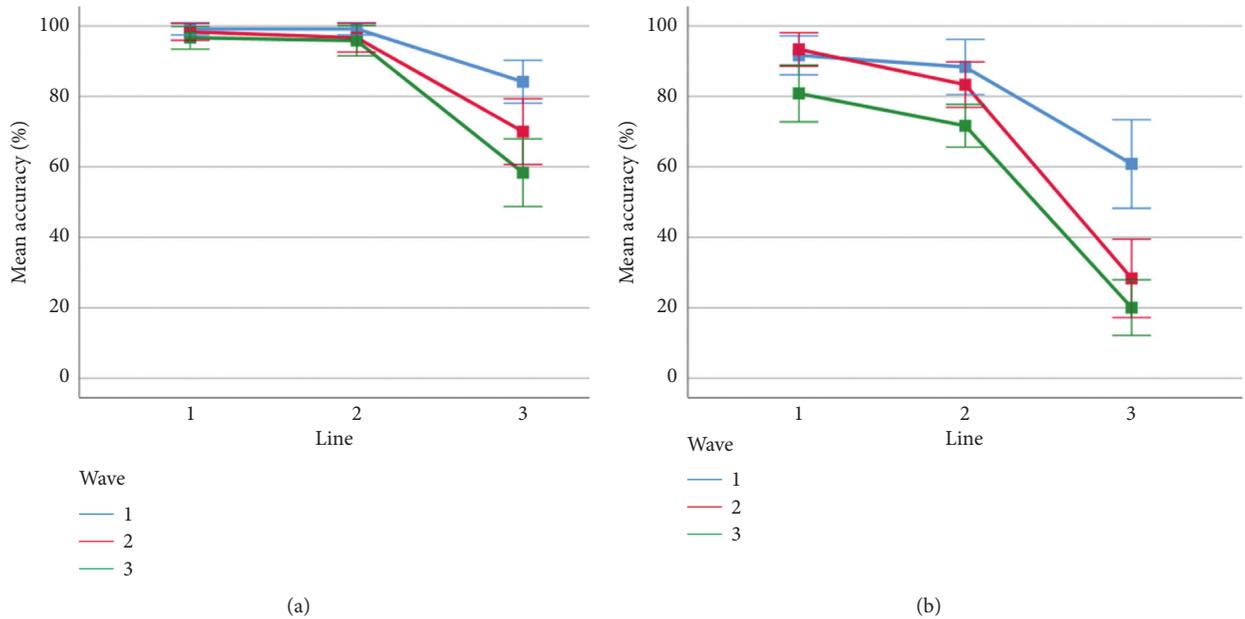

(a)

(b)

Figure 6: Average accuracy percentages observed for the levels of line and wave when rotation was in off (a) and on (b) conditions. Error bars indicate %95 CI.

fNIRS data epochs for the rest and task periods were extracted from the continuous data based on time synchronization markers. The markers were recorded by COBI Studio during the experiment. During the split operation, technical problems have been detected related to markers on 13 out of 450 main blocks, possibly due to the RS232 buffer limitations. These problematic markers were corrected with the help of the behavioral data logged by the web application. Blood oxygenation changes within each optode were calculated using the modified Beer-Lambert law with reference to rest periods at the beginning of each trial with fNIRSoft. This process provided four different measures for each block, namely, relative changes in oxyhemoglobin (HbO), deoxyhemoglobin (HbR), total hemoglobin (i.e., HbO + HbR, abbreviated HbT), and oxygenation (i.e., HbO–HbR, abbreviated as Oxy), which were used as features for our subsequent investigation of the participants' brain responses to different CAPTCHAs. The block averages for each measure were then computed and consolidated into a single file along with CAPTCHA types and participant information. Finally, checks for parametric assumptions and outliers resulted in eliminating three more participants from the sample. The final data set included block averages of 22 participants over 18 CAPTCHA combinations.

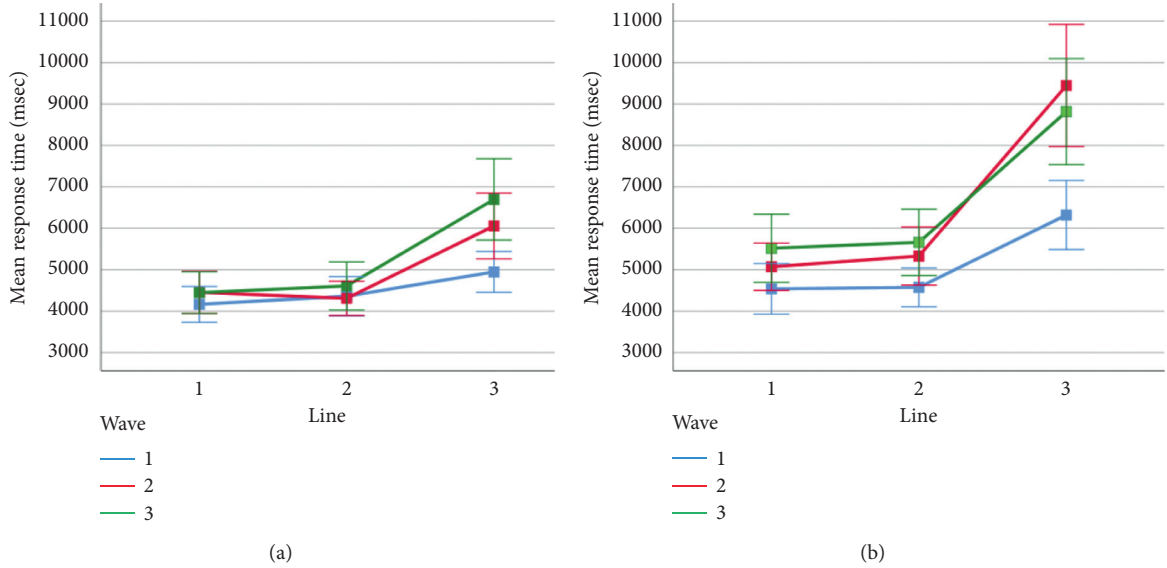

Figure 7: Average response times observed for the levels of line and wave when rotation was in off (a) and on (b) conditions. Error bars indicate %95 CI.

Table 3: Detected significant changes in mean values (tests of within-subjects effects).

| Method | | | | Type III SS | df | MS | F | p | Partial $\eta 2$ |
|---|---|---|---|---|---|---|---|---|---|
| HbO | o1 | Line ∗ wave ∗ rotation | Sphericity assumed | 1.657 | 4 | 0.414 | 3.892 | 0.006 | 0.170 |
| | o11 | Line ∗ wave ∗ rotation | Sphericity assumed | 1.410 | 4 | 0.352 | 3.427 | 0.012 | 0.146 |
| | o12 | Line ∗ wave ∗ rotation | Sphericity assumed | 1.044 | 4 | 0.261 | 2.587 | 0.044 | 0.126 |
| | o13 | Line ∗ wave ∗ rotation | Sphericity assumed | 1.039 | 4 | 0.260 | 2.554 | 0.046 | 0.118 |
| | o16 | Wave ∗ rotation | Sphericity assumed | 0.742 | 2 | 0.371 | 3.686 | 0.034 | 0.156 |
| | | Line ∗ wave ∗ rotation | Sphericity assumed | 1.092 | 4 | 0.273 | 3.176 | 0.018 | 0.137 |
| | o2 | Line ∗ wave ∗ rotation | Sphericity assumed | 1.203 | 4 | 0.301 | 2.736 | 0.034 | 0.120 |
| | o3 | Line ∗ wave ∗ rotation | Sphericity assumed | 1.115 | 4 | 0.279 | 2.588 | 0.043 | 0.115 |
| | o4 | Line ∗ wave ∗ rotation | Sphericity assumed | 1.603 | 4 | 0.401 | 3.730 | 0.008 | 0.172 |
| HbR | o6 | Wave ∗ rotation | Sphericity assumed | 0.206 | 2 | 0.103 | 3.712 | 0.034 | 0.171 |
| HbT | o1 | Line ∗ wave ∗ rotation | Sphericity assumed | 1.665 | 4 | 0.416 | 4.024 | 0.005 | 0.175 |
| | o10 | Line ∗ rotation | Sphericity assumed | 2.098 | 2 | 1.049 | 4.180 | 0.023 | 0.180 |
| | o12 | Line ∗ wave ∗ rotation | Sphericity assumed | 1.156 | 4 | 0.289 | 2.542 | 0.047 | 0.124 |
| | o14 | Line ∗ wave ∗ rotation | Sphericity assumed | 1.172 | 4 | 0.293 | 2.664 | 0.040 | 0.143 |
| | o16 | Line ∗ wave ∗ rotation | Sphericity assumed | 1.091 | 4 | 0.273 | 3.185 | 0.018 | 0.137 |
| | o2 | Line ∗ wave ∗ rotation | Sphericity assumed | 1.150 | 4 | 0.287 | 2.624 | 0.041 | 0.116 |
| | o4 | Line ∗ wave ∗ rotation | Sphericity assumed | 1.285 | 4 | 0.321 | 2.663 | 0.039 | 0.129 |
| | o8 | Line ∗ rotation | Sphericity assumed | 2.027 | 2 | 1.013 | 3.586 | 0.038 | 0.166 |
| Oxy | o1 | Line ∗ wave ∗ rotation | Sphericity assumed | 1.827 | 4 | 0.457 | 3.425 | 0.013 | 0.153 |
| | o10 | Line ∗ wave ∗ rotation | Sphericity assumed | 1.517 | 4 | 0.379 | 2.850 | 0.029 | 0.130 |
| | o11 | Line ∗ wave ∗ rotation | Sphericity assumed | 2.120 | 4 | 0.530 | 3.554 | 0.010 | 0.151 |
| | o13 | Line ∗ wave ∗ rotation | Sphericity assumed | 1.631 | 4 | 0.408 | 2.533 | 0.047 | 0.118 |
| | o3 | Line ∗ wave ∗ rotation | Sphericity assumed | 2.029 | 4 | 0.507 | 3.022 | 0.022 | 0.131 |
| | o4 | Line ∗ wave ∗ rotation | Sphericity assumed | 2.004 | 4 | 0.501 | 4.108 | 0.005 | 0.186 |
| | o9 | Line ∗ wave ∗ rotation | Sphericity assumed | 1.884 | 4 | 0.471 | 3.464 | 0.012 | 0.148 |

After the preprocessing, a repeated-measures ANOVA was conducted to analyze the relationship between visual features and fNIRS variables. Groups were created according to optodes (o1–o16) and dependent variables (HbO, HbR, HbT, and Oxy). The full results of the analysis are presented in Tables 3 and 4. The temporal mean plots for the HbO and

TABLE 4: Significant contrast results following the three-way interaction between line, wave, and rotation for HbO measurements.

| Optode | Source | Line | Wave | Rotation | Type III SS | df | MS | F | p | Partial $\eta^2$ |
|---|---|---|---|---|---|---|---|---|---|---|
| 1 | Line * wave * rotation | Linear | Linear | Linear | 0.568 | 1 | 0.568 | 5.428 | 0.031 | 0.222 |
| | | Quadratic | Linear | Linear | 0.833 | 1 | 0.833 | 6.149 | 0.023 | 0.245 |
| | Error | Linear | Linear | Linear | 1.989 | 19 | 0.105 | | | |
| | | Quadratic | Linear | Linear | 2.573 | 19 | 0.135 | | | |
| 2 | Line * wave * rotation | Linear | Linear | Linear | 0.568 | 1 | 0.568 | 4.293 | 0.051 | 0.177 |
| | Error | Linear | Linear | Linear | 2.646 | 20 | 0.132 | | | |
| 3 | Line * wave * rotation | Linear | Linear | Linear | 0.466 | 1 | 0.466 | 5.909 | 0.025 | 0.228 |
| | Error | Linear | Linear | Linear | 1.576 | 20 | 0.079 | | | |
| 4 | Line * wave * rotation | Linear | Linear | Linear | 0.584 | 1 | 0.584 | 4.374 | 0.051 | 0.196 |
| | | Quadratic | Linear | Linear | 0.688 | 1 | 0.688 | 5.139 | 0.036 | 0.222 |
| | Error | Linear | Linear | Linear | 2.404 | 18 | 0.134 | | | |
| | | Quadratic | Linear | Linear | 2.411 | 18 | 0.134 | | | |
| 9 | Line * wave * rotation | Quadratic | Linear | Linear | 0.711 | 1 | 0.711 | 6.563 | 0.019 | 0.247 |
| | Error | Quadratic | Linear | Linear | 2.167 | 20 | 0.108 | | | |
| 10 | Line * wave * rotation | Quadratic | Linear | Linear | 0.763 | 1 | 0.763 | 4.607 | 0.045 | 0.195 |
| | Error | Quadratic | Linear | Linear | 3.145 | 19 | 0.166 | | | |
| 11 | Line * wave * rotation | Linear | Linear | Linear | 0.534 | 1 | 0.534 | 5.34 | 0.032 | 0.211 |
| | | Quadratic | Linear | Linear | 0.627 | 1 | 0.627 | 4.794 | 0.041 | 0.193 |
| | Error | Linear | Linear | Linear | 1.999 | 20 | 0.1 | | | |
| | | Quadratic | Linear | Linear | 2.617 | 20 | 0.131 | | | |
| 12 | Line * wave * rotation | Linear | Linear | Linear | 0.543 | 1 | 0.543 | 6.202 | 0.023 | 0.256 |
| | Error | Linear | Linear | Linear | 1.576 | 18 | 0.088 | | | |
| 13 | Line * wave * rotation | Linear | Linear | Linear | 0.621 | 1 | 0.621 | 5.366 | 0.032 | 0.22 |
| | Error | Linear | Linear | Linear | 2.198 | 19 | 0.116 | | | |
| 14 | Line * wave * rotation | Linear | Linear | Linear | 0.965 | 1 | 0.965 | 8.218 | 0.011 | 0.339 |
| | Error | Linear | Linear | Linear | 1.878 | 16 | 0.117 | | | |
| 15 | Line * wave * rotation | Linear | Linear | Linear | 0.676 | 1 | 0.676 | 8.225 | 0.012 | 0.354 |
| | Error | Linear | Linear | Linear | 1.232 | 15 | 0.082 | | | |
| 16 | Line * wave * rotation | Linear | Linear | Linear | 0.825 | 1 | 0.825 | 8.038 | 0.010 | 0.287 |
| | | Quadratic | Linear | Linear | 0.23 | 1 | 0.23 | 3.652 | 0.070 | 0.154 |
| | Error | Linear | Linear | Linear | 2.052 | 20 | 0.103 | | | |
| | | Quadratic | Linear | Linear | 1.26 | 20 | 0.063 | | | |

HbR responses observed at each optode for all stimuli types are also presented in Appendix. Here, we present a summary of the findings. There was no observed main effect that returned a significant difference for any of the HbO, HbR, HbT, or Oxy measurements. However, we obtained significant interaction effects. We report the results for HbO since they are expected to provide the largest modulation in the fNIRS signals due to the way the vascular system responds to oxygen demands of active neural tissue. The results showed a statistically significant three-way interaction between line, wave, and rotation on optode1, optode2, optode3, optode4, optode11, optode12, optode13, optode16. Figure 8 shows the projection of F-ratios corresponding to the three-way interaction effects observed at each optode over the prefrontal cortex. The visualization is produced with a B-spline interpolation in fNIR Soft [30] to show the regions where we observed significant three-way interaction at the $\alpha$ level of .05. The most significant interaction in average HbO values is clustered around bilateral dorsolateral PFC and regions in right dorsomedial PFC, consistent with implicated regions in related neuroimaging studies focusing on neural correlates of visuospatial and orthographic processing tasks with varying difficulty levels [33]. Planned contrasts following the three-way ANOVA found significant trends that linearly and quadratically differ across levels of visual features employed. The results are summarized in Table 4 together with mean plots. The trends suggest that the interaction effect is mainly due to the drop in oxygenation during the transition from thin to thick lines when the rotation and the last level of wave distortion were in effect. When viewed together with the sudden decrease in accuracy measures, this may suggest that the CAPTCHAs turned out to be too difficult to resolve when all distortions were applied together, so the participants may have switched to a guessing strategy, relieving the attentional resources associated with the frontoparietal networks.

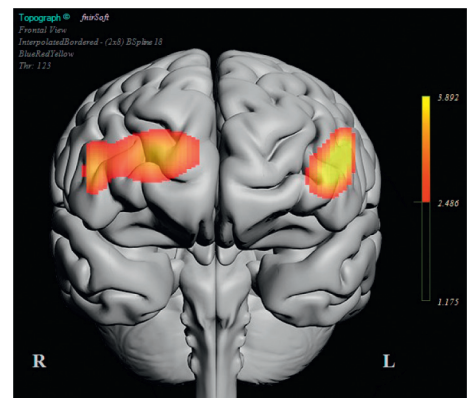

FIGURE 8: A B-spline interpolated projection of F-ratios corresponding to the three-way interaction effects observed for the HbO changes observed over the PFC.

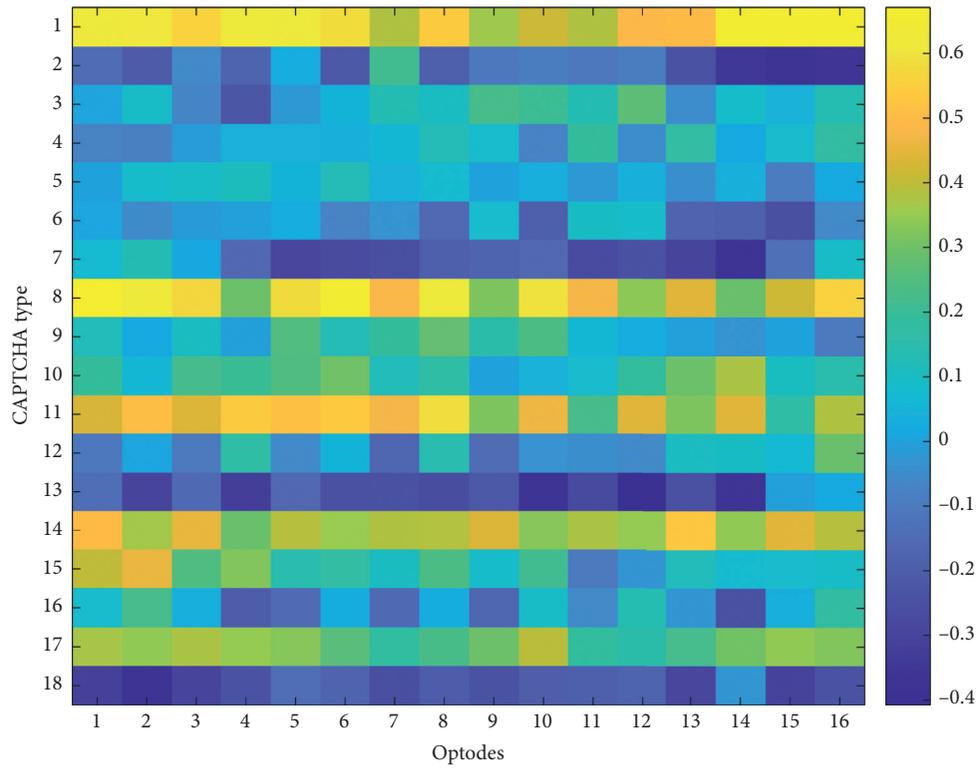

Figure 9: Pearson correlation coefficients between average accuracy and average HbO changes observed at 16 different optodes for each CAPTCHA block type.

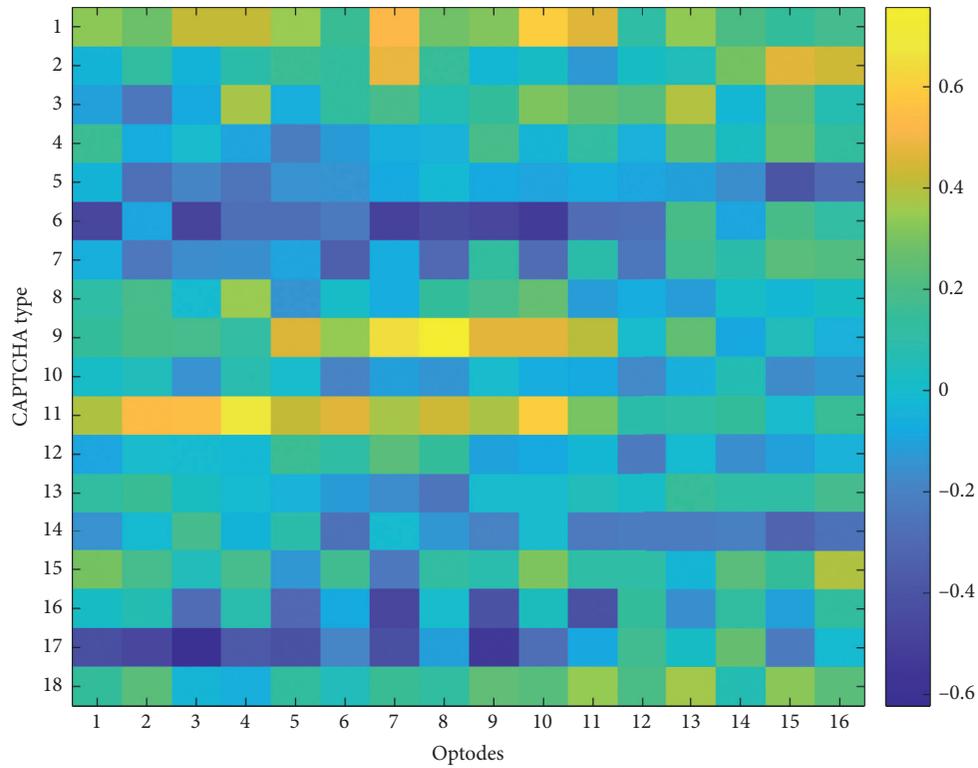

Figure 10: Pearson correlation coefficients between average accuracy and average HbR changes observed at 16 different optodes for each CAPTCHA block type.

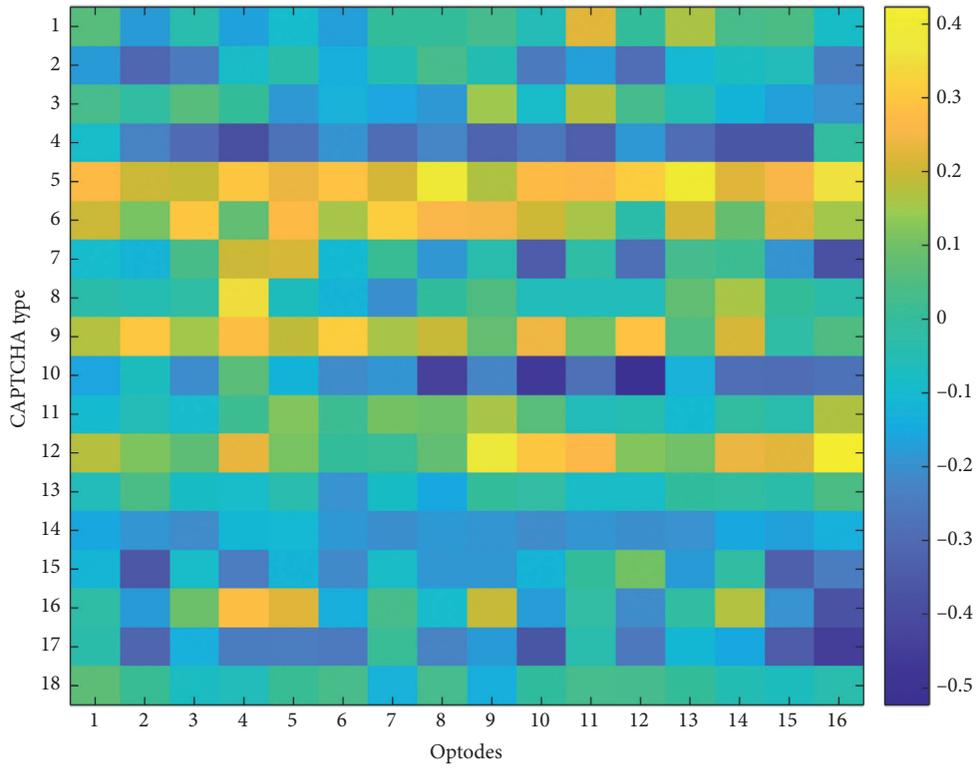

Figure 11: Pearson correlation coefficients between average response times and average HbO changes observed at 16 different optodes for each CAPTCHA block type.

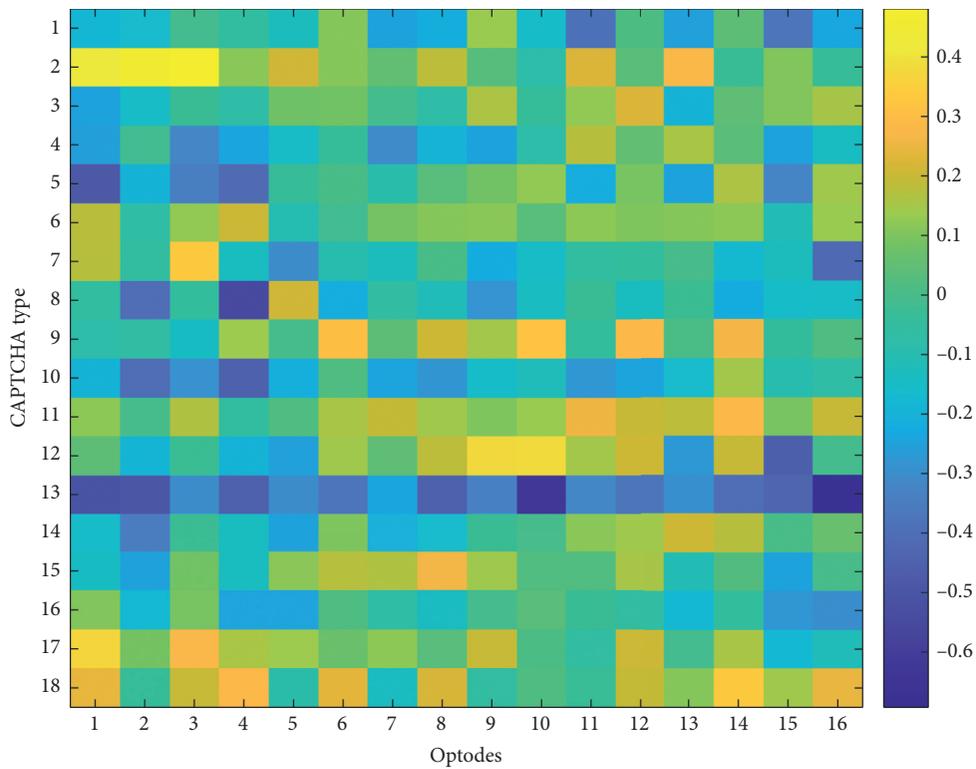

Figure 12: Pearson correlation coefficients between average response times and average HbR changes observed at 16 different optodes for each CAPTCHA block type.

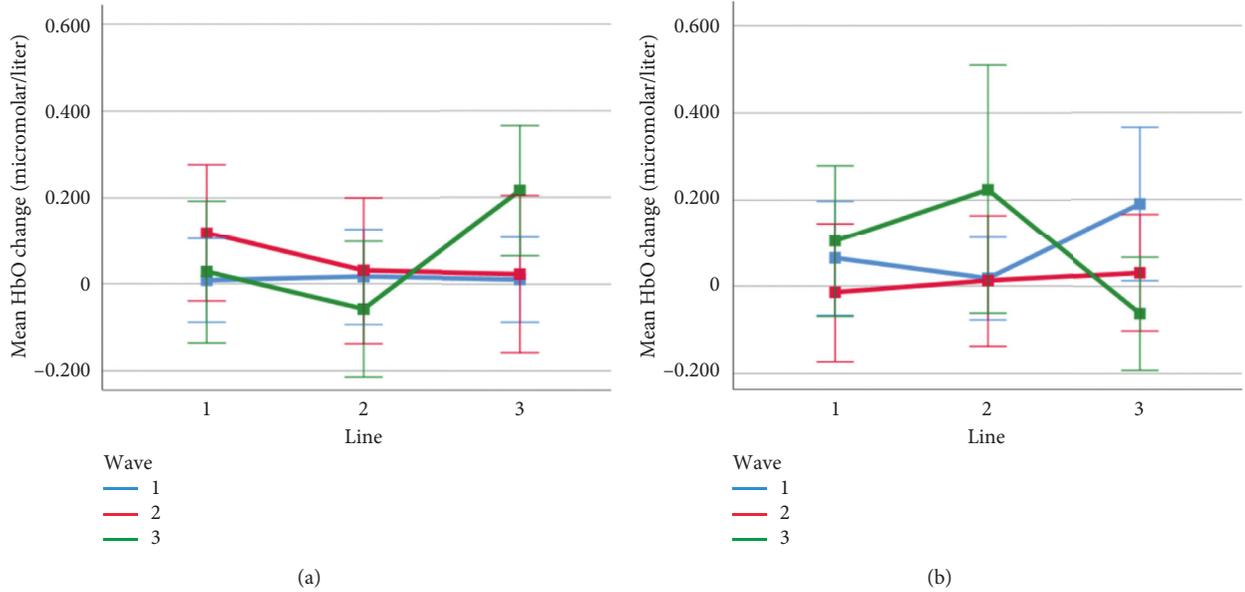

Figure 13: Mean HbO change observed for line and wave conditions at optode 1 for rotation 1 (a) and rotation 2 (b) cases. Error bars indicate %95 CI.

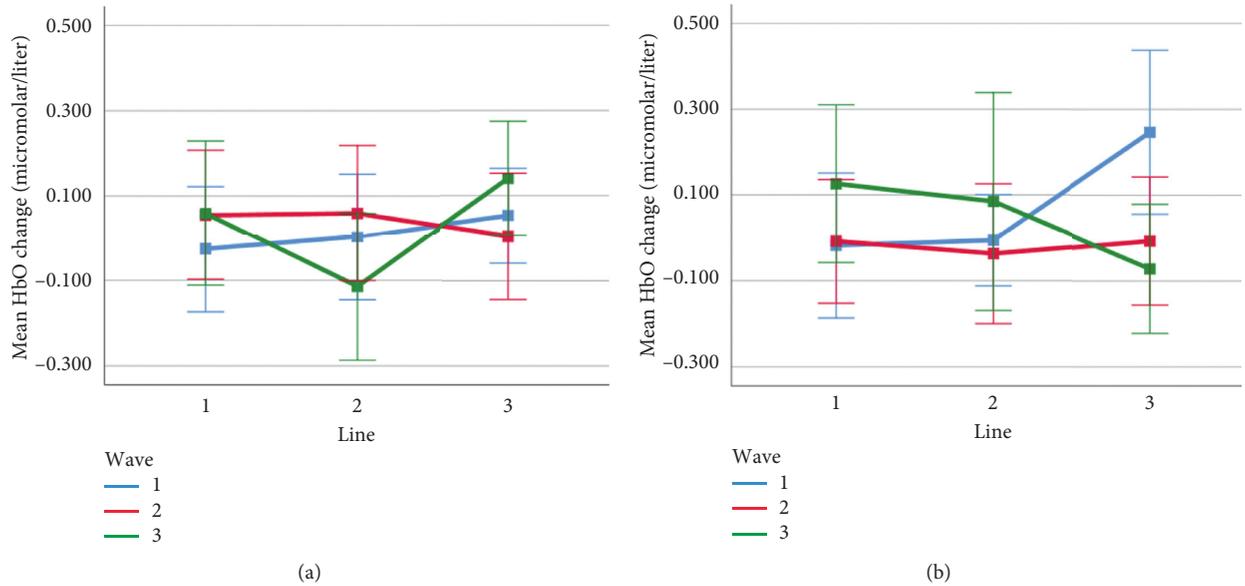

Figure 14: Mean HbO change observed for line and wave conditions at optode 2 for rotation 1 (a) and rotation 2 (b) cases. Error bars indicate %95 CI.

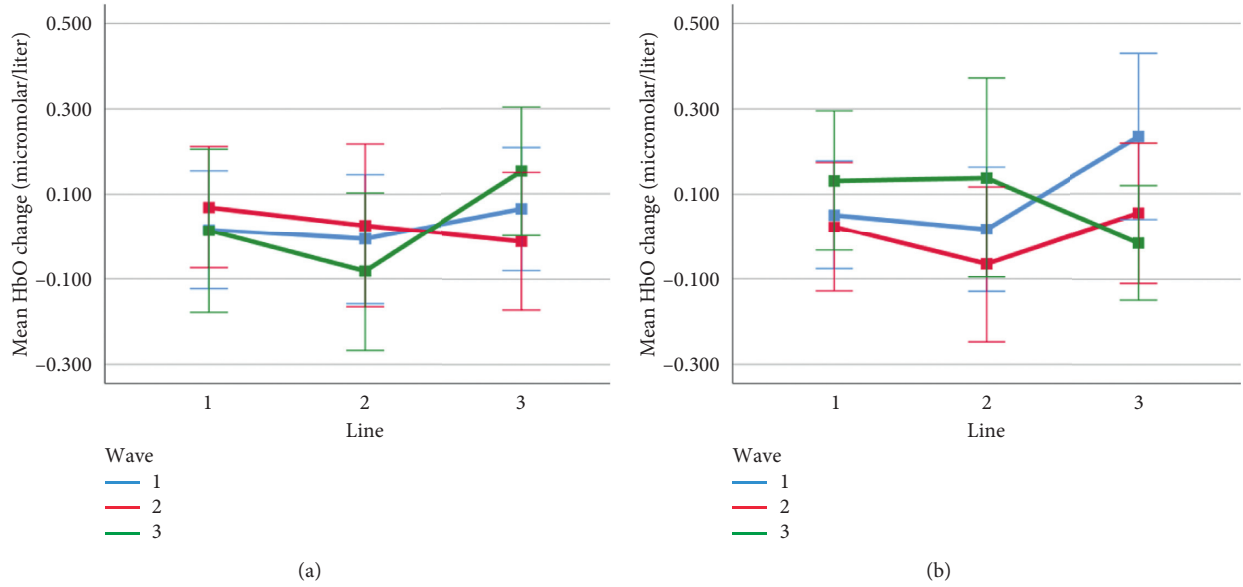

Figure 15: Mean HbO change observed for line and wave conditions at optode 3 for rotation 1 (a) and rotation 2 (b) cases. Error bars indicate %95 CI.

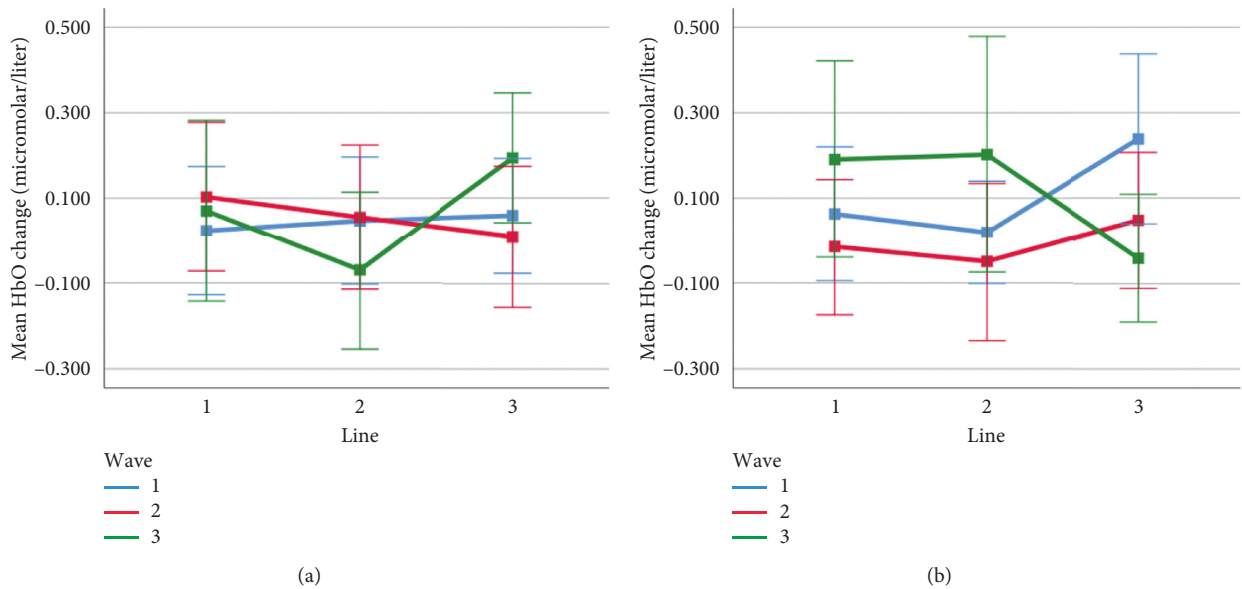

Figure 16: Mean HbO change observed for line and wave conditions at optode 4 for rotation 1 (a) and rotation 2 (b) cases. Error bars indicate %95 CI.

3.3. Relationship between Behavioral and Neuroimaging Results. To explore the relationship between optical imaging and behavioral performance measures, Pearson *r* correlation coefficients were computed among block averages for HbO and HbR changes observed at each optode and the average accuracy and the average response time for each CAPTCHA block. The accuracy and response time results are summarized by the heatmaps in Figures 9–12, respectively. The

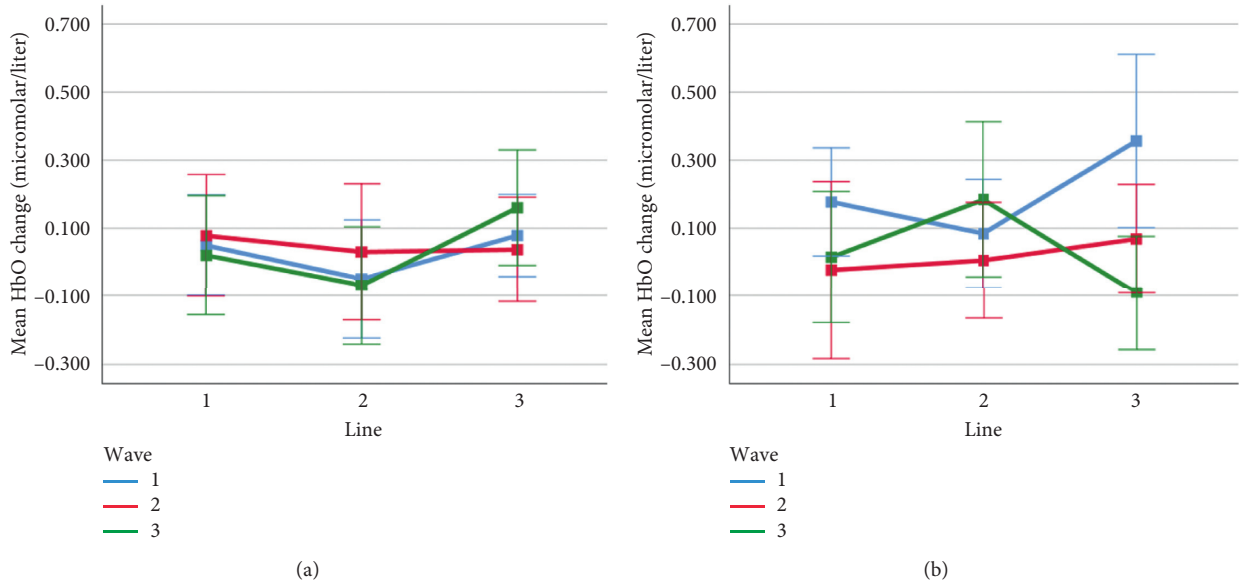

Figure 17: Mean HbO change observed for line and wave conditions at optode 9 for rotation 1 (a) and rotation 2 (b) cases. Error bars indicate %95 CI.

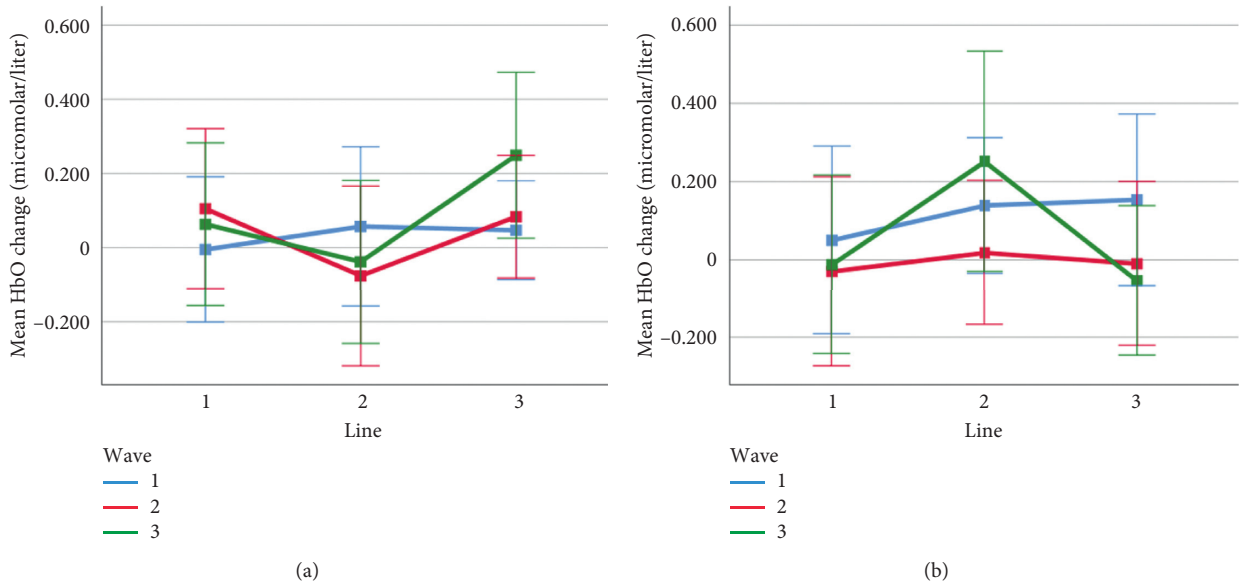

Figure 18: Mean HbO change observed for line and wave conditions at optode 10 for rotation 1 (a) and rotation 2 (b) cases. Error bars indicate %95 CI.

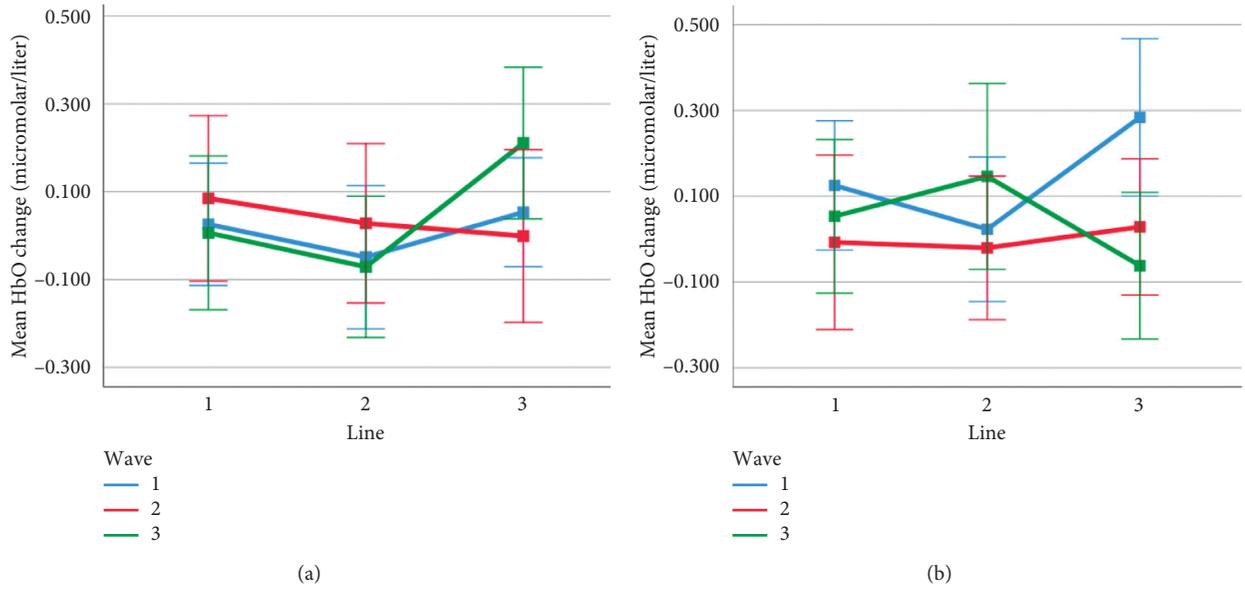

Figure 19: Mean HbO change observed for line and wave conditions at optode 11 for rotation 1 (a) and rotation 2 (b) cases. Error bars indicate %95 CI.

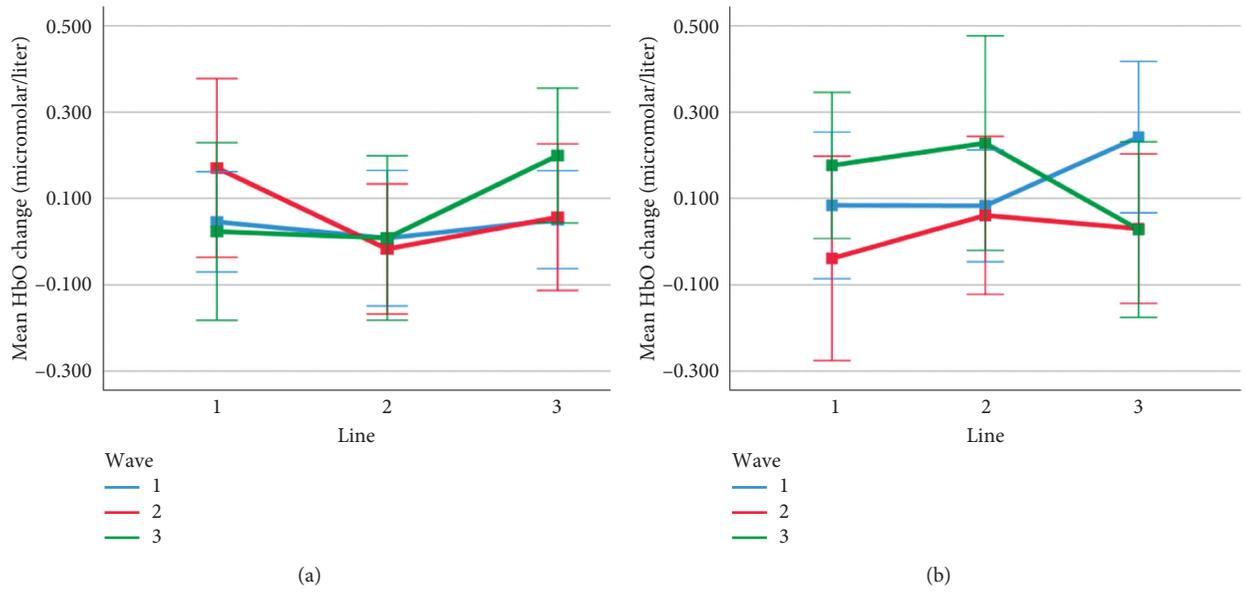

Figure 20: Mean HbO change observed for line and wave conditions at optode 12 for rotation 1 (a) and rotation 2 (b) cases. Error bars indicate %95 CI.

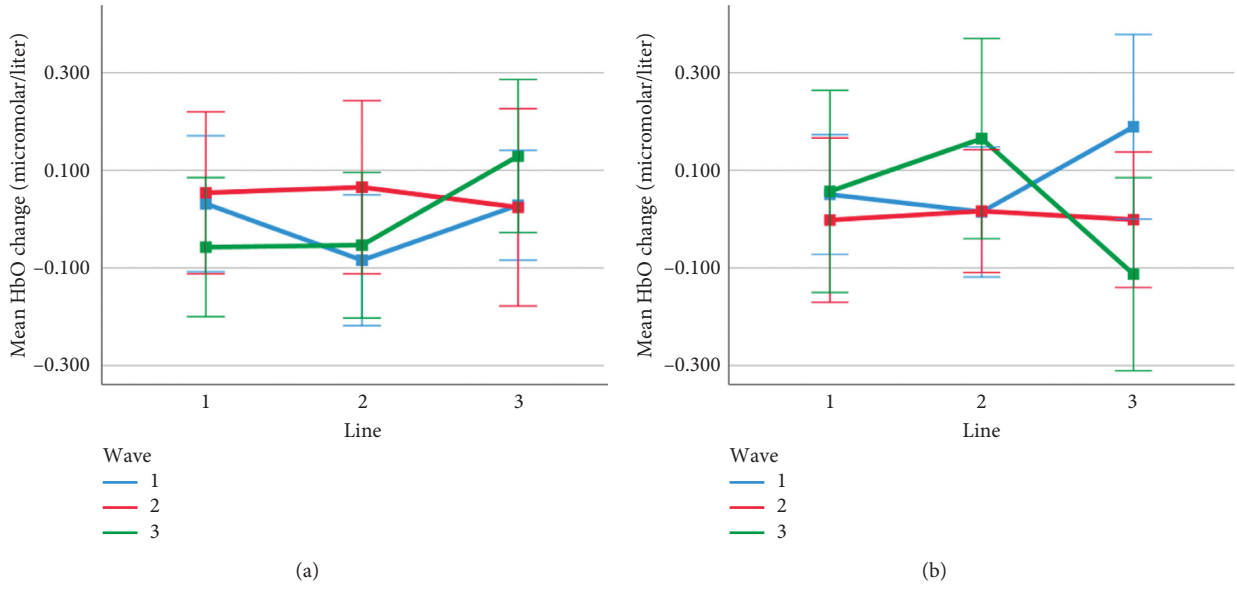

Figure 21: Mean HbO change observed for line and wave conditions at optode 13 for rotation 1 (a) and rotation 2 (b) cases. Error bars indicate %95 CI.

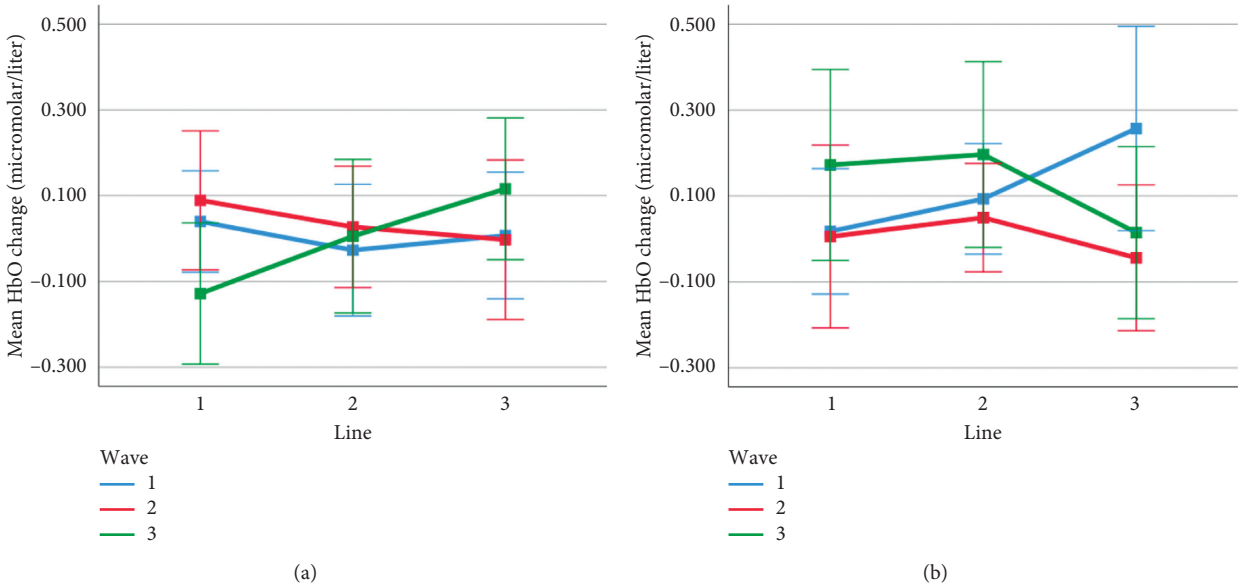

Figure 22: Mean HbO change observed for line and wave conditions at optode 14 for rotation 1 (a) and rotation 2 (b) cases. Error bars indicate %95 CI.

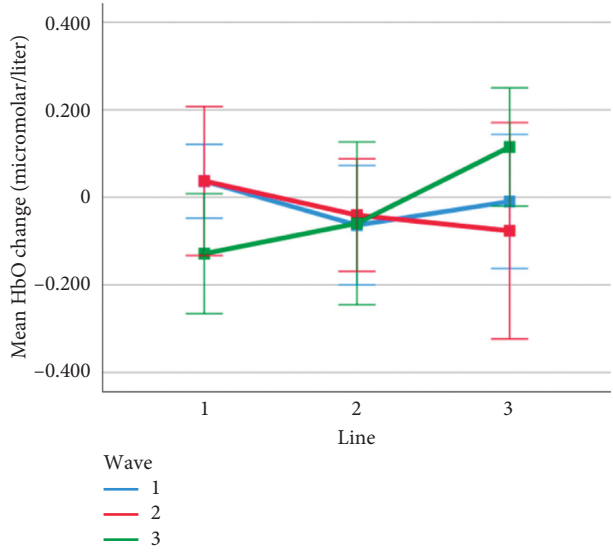

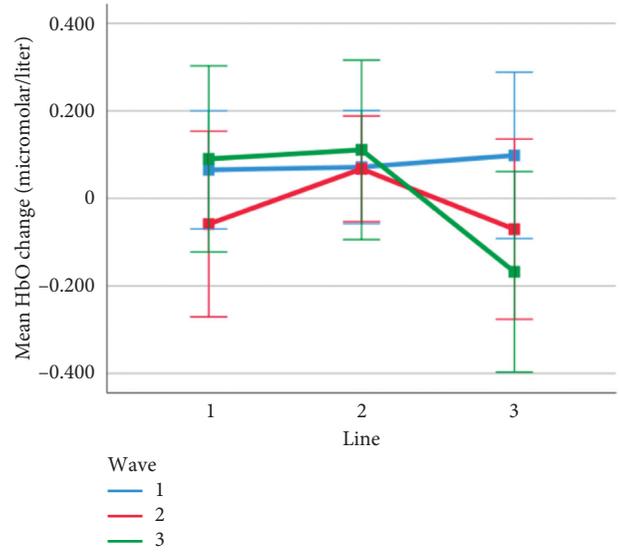

(a)

(b)

Figure 23: Mean HbO change observed for line and wave conditions at optode 15 for rotation 1 (a) and rotation 2 (b) cases. Error bars indicate %95 CI.

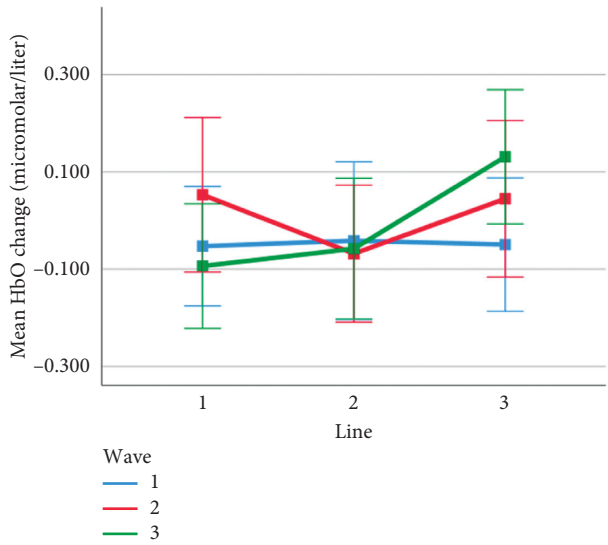

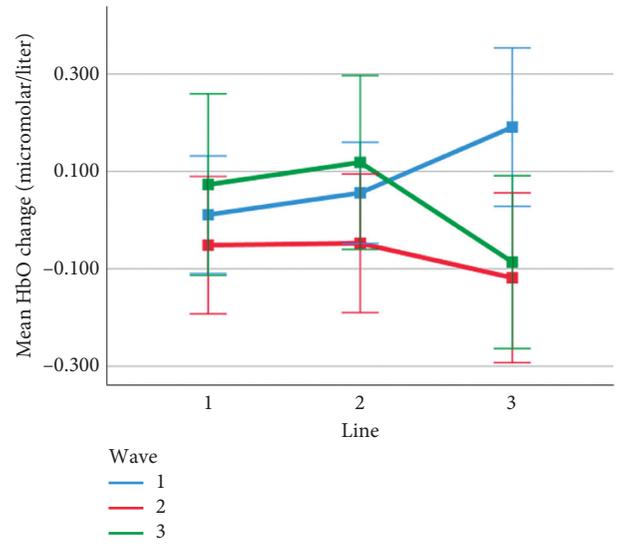

(a)

(b)

Figure 24: Mean HbO change observed for line and wave conditions at optode 16 for rotation 1 (a) and rotation 2 (b) cases. Error bars indicate %95 CI.

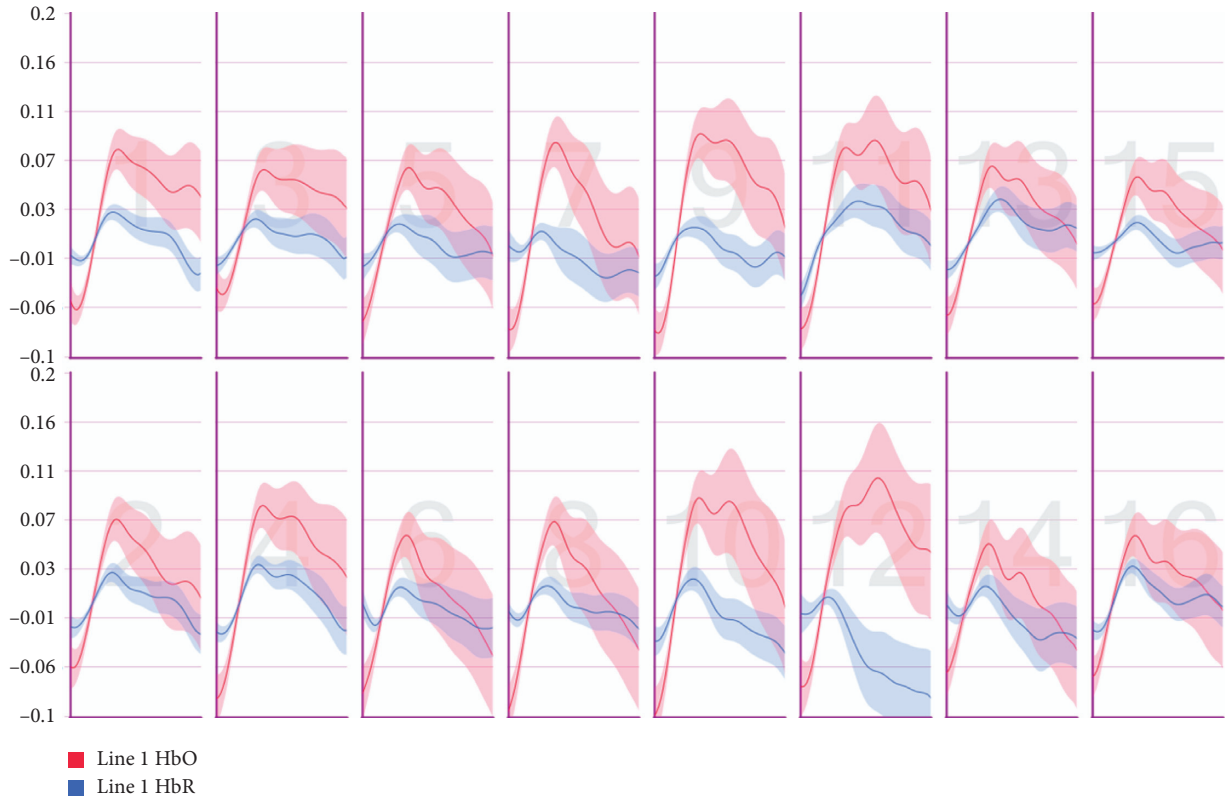

FIGURE 25: Temporal HbO and HbR averages observed for the first 30 seconds of CAPTCHA blocks with difficulty level 1 for the line feature.

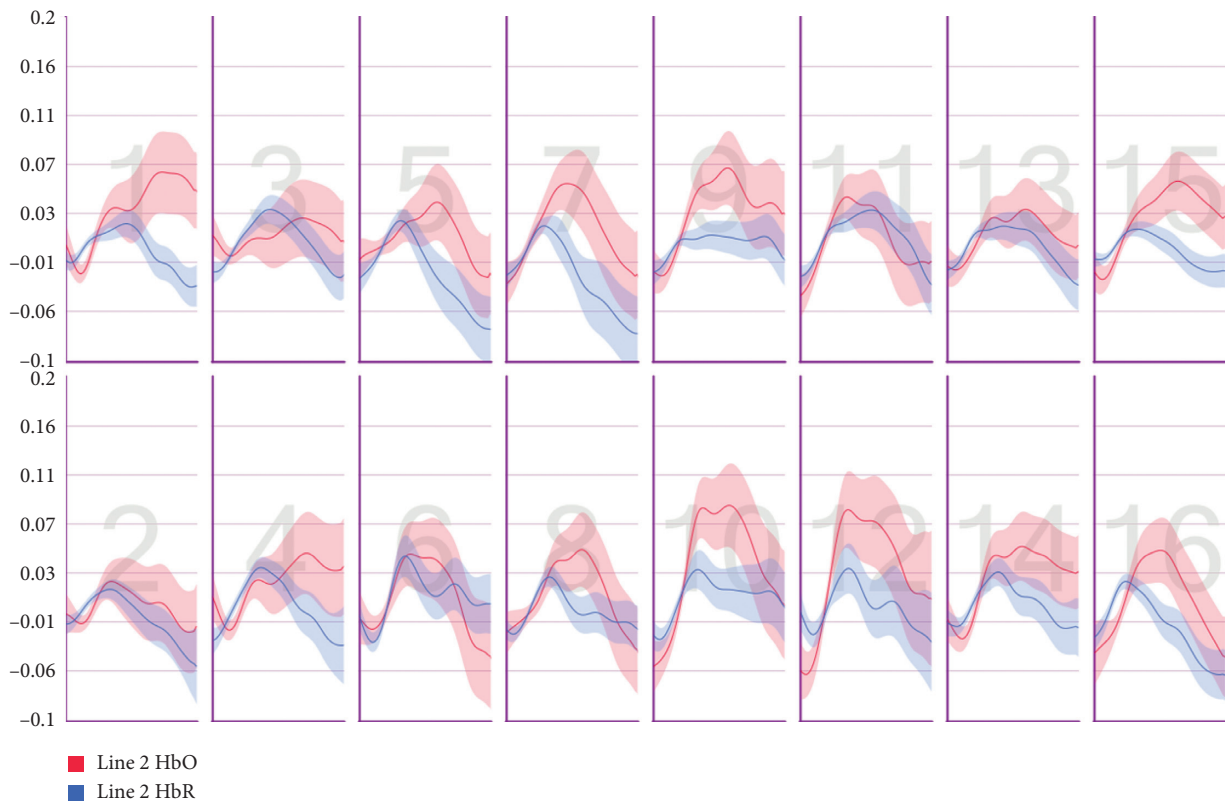

FIGURE 26: Temporal HbO and HbR averages observed for the first 30 seconds of CAPTCHA blocks with difficulty level 2 for the line feature.

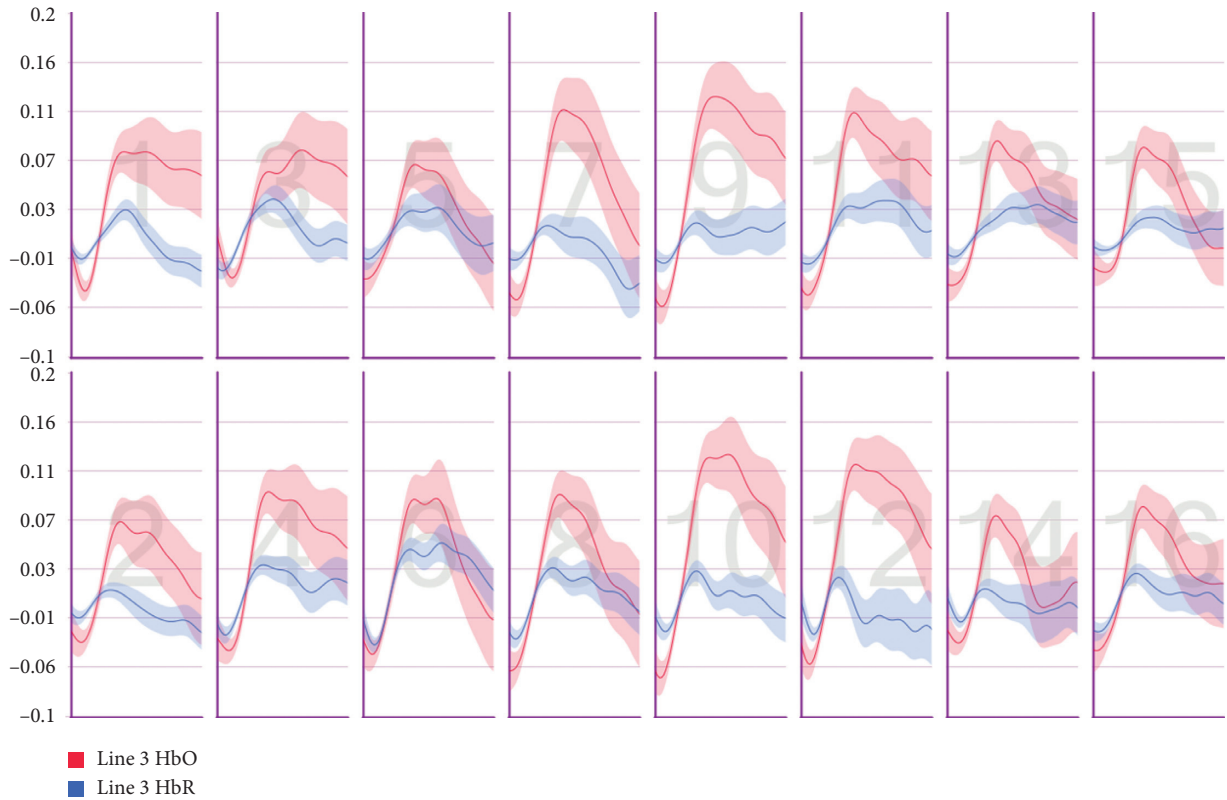

Figure 27: Temporal HbO and HbR averages observed for the first 30 seconds of CAPTCHA blocks with difficulty level 3 for the line feature.

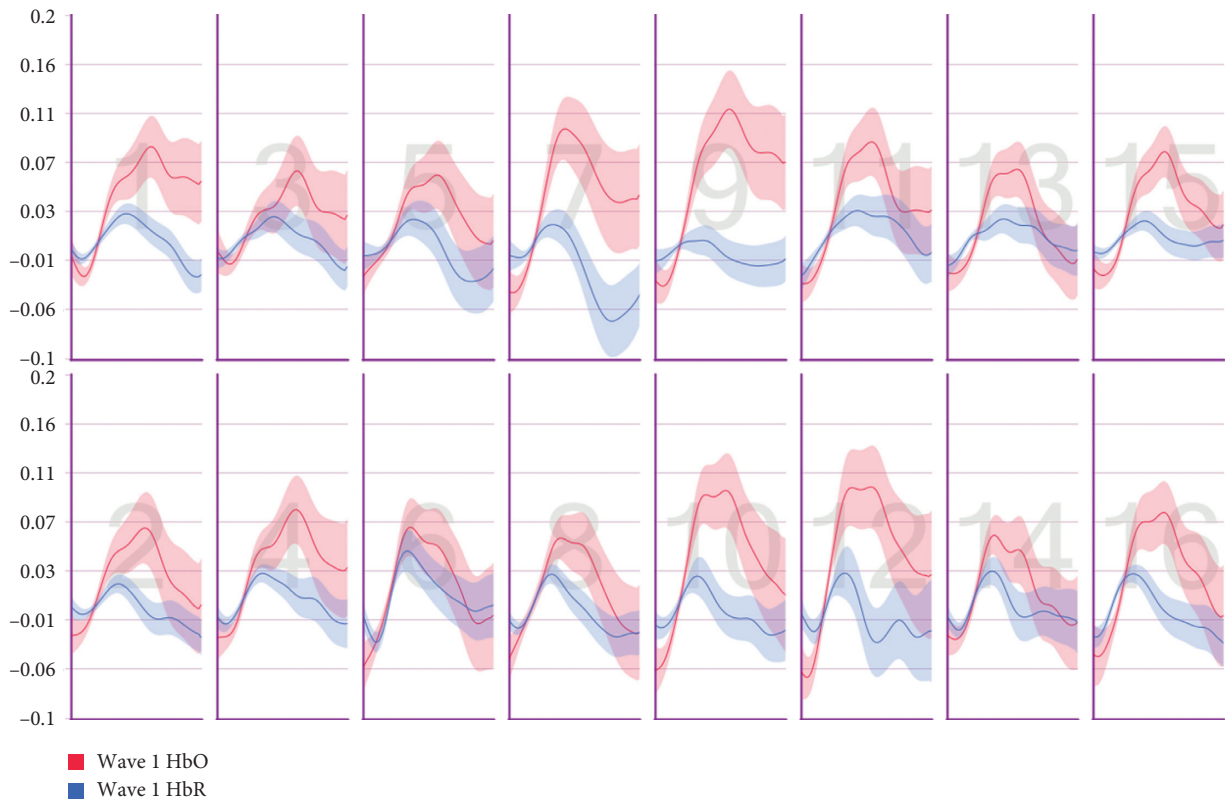

Figure 28: Temporal HbO and HbR averages observed for the first 30 seconds of CAPTCHA blocks with difficulty level 1 for the wave feature.

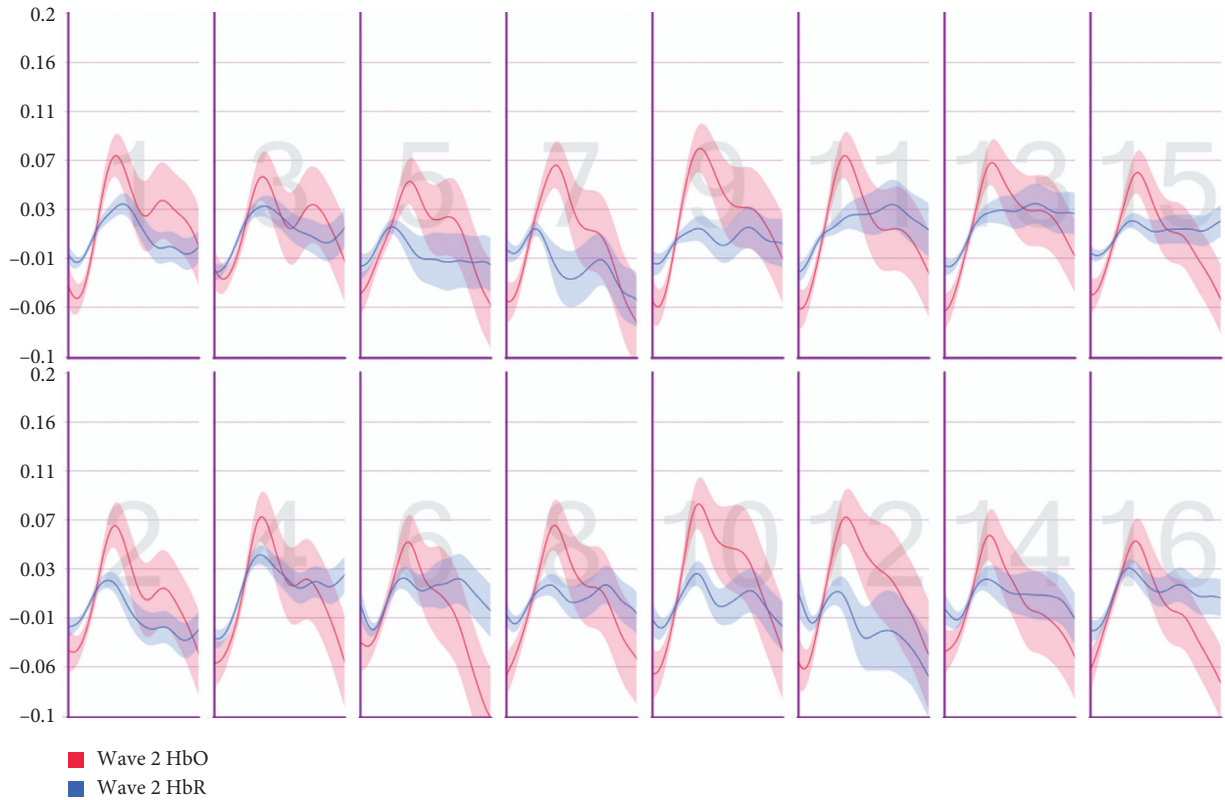

FIGURE 29: Temporal HbO and HbR averages observed for the first 30 seconds of CAPTCHA blocks with difficulty level 2 for the wave feature.

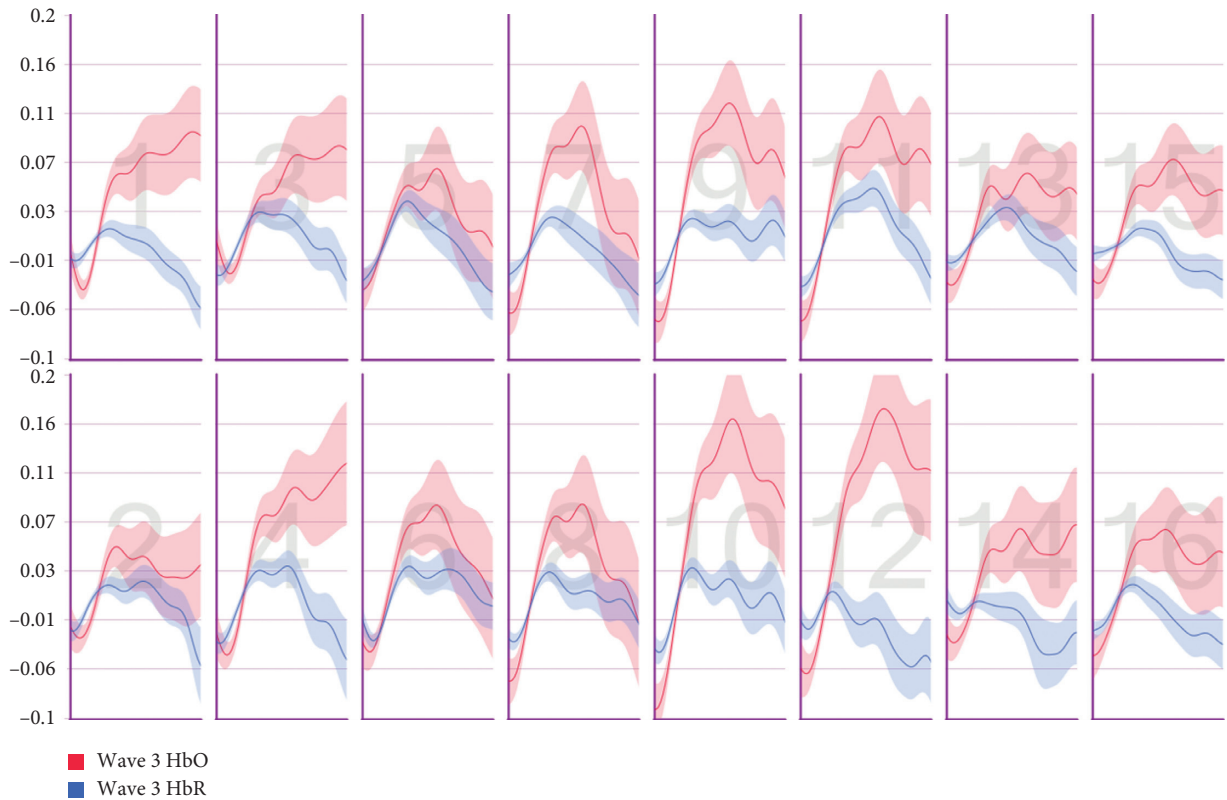

FIGURE 30: Temporal HbO and HbR averages observed for the first 30 seconds of CAPTCHA blocks with difficulty level 3 for the wave feature.

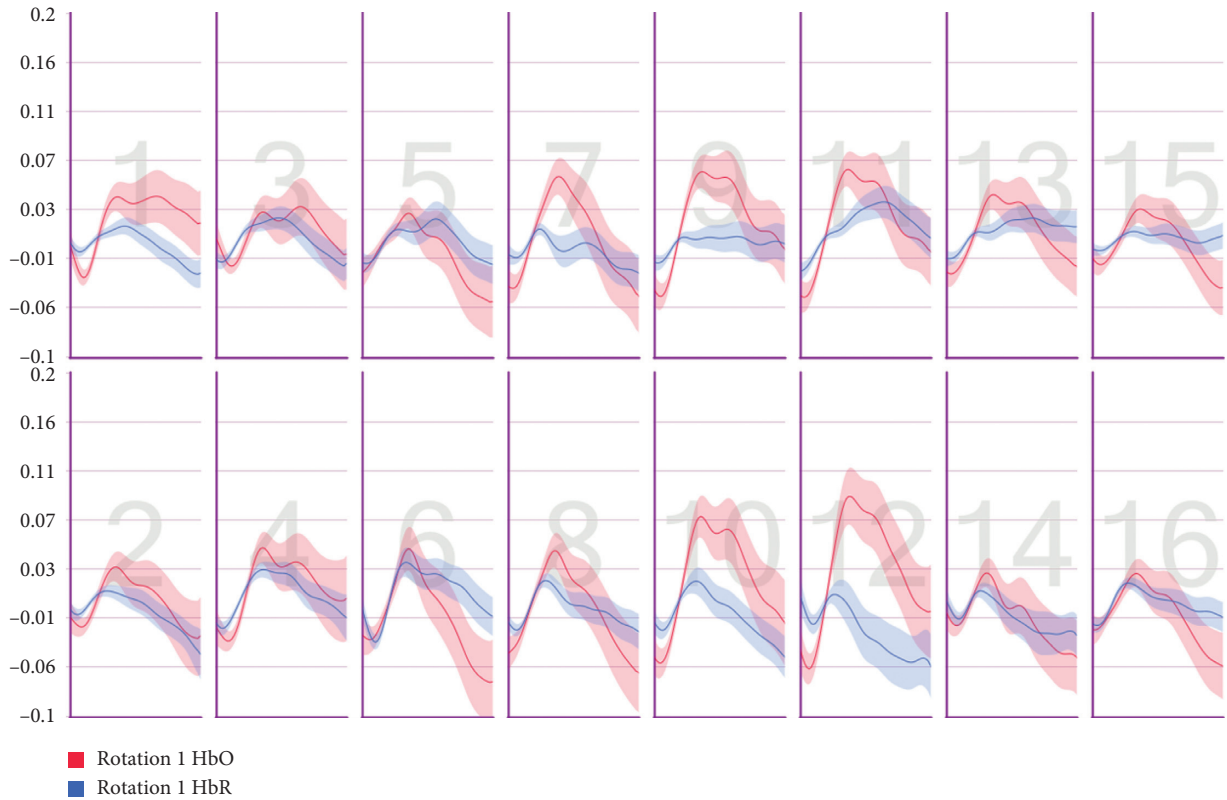

Figure 31: Temporal HbO and HbR averages observed for the first 30 seconds of CAPTCHA blocks with difficulty level 1 for the rotation feature.

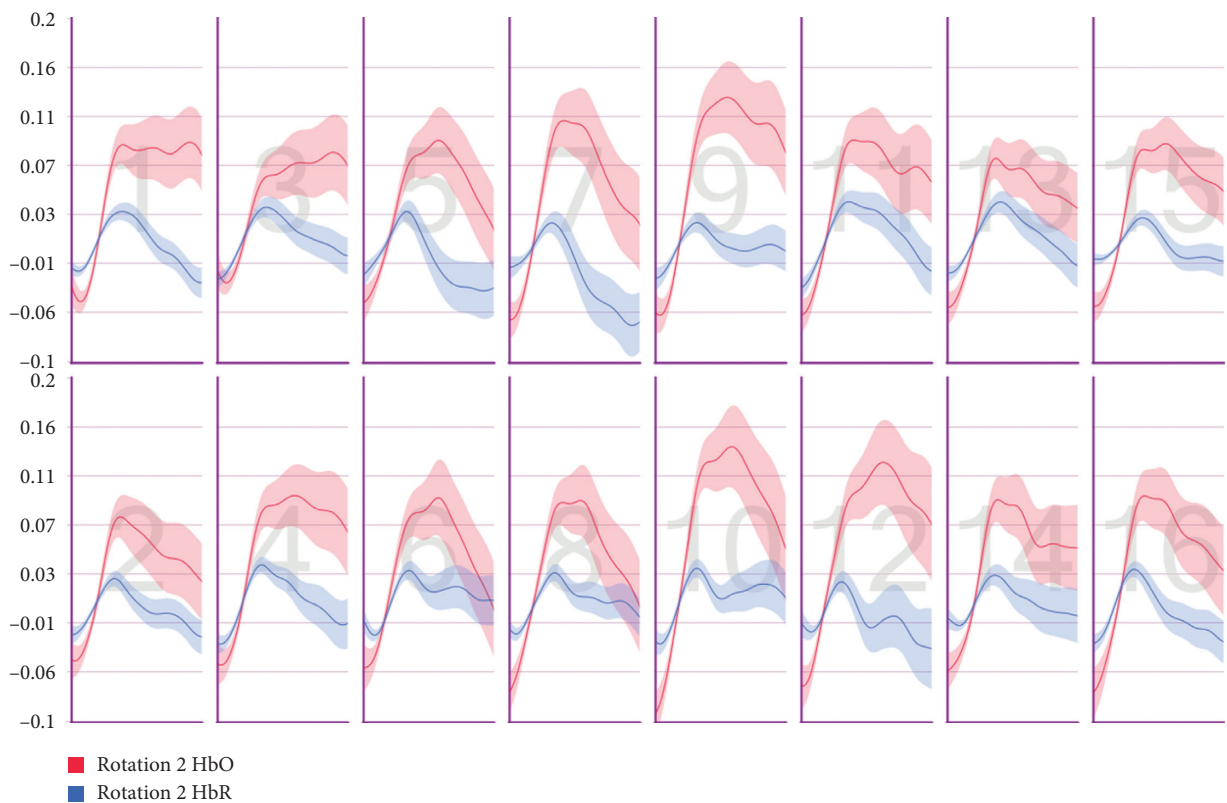

Figure 32: Temporal HbO and HbR averages observed for the first 30 seconds of CAPTCHA blocks with difficulty level 2 for the rotation feature.

maps highlight those conditions that exhibit a linear relationship between behavioral performance and HbO/HbR changes. The quadratic shifts in HbO trends observed as distortions were combined in subsequent CAPTCHA types seemed to break the linearity in the relationship between behavioral and fNIRS results.

## 4. Conclusion

The purpose of the present study was to investigate whether a text-based CAPTCHA solving task would produce a differentiable, hemodynamic response in the human prefrontal cortex, as measured by a portable fNIRS device, as a function of three visual features (line, wave, and rotation). The features were selected based on the previous research reported by Google reCAPTCHA v1 final design [16]. Eighteen types of CAPTCHA were presented to the participants as combinations of the selected features with different difficulty levels. We expected that the resultx of the current study might contribute methodologically to cybersecurity, neuroscience, and usability designers in CAPTCHA designs and the design of similar challenges for authentication.

The behavioral results were primarily compatible with the previous findings in the literature. A specific contribution of our study is that although all selected features, the line, wave, and rotation, returned statistically significant findings in performance, their combined effects had a nonadditive impact on behavioral performance indicators, given the significantly larger reduction in accuracy and increase in response time when stimulus switched from thin to thick line with some wave and rotation distortions. The fNIRS results revealed statistically significant interaction effects, which showed that the selected features do not exhibit linear, additive effects. When viewed together with the sudden decrease in accuracy measures, this may suggest that the CAPTCHAs became too challenging to resolve when all distortions were applied together, so the participants may have switched to a guessing strategy, relieving the attentional resources associated with the frontoparietal networks. It is also likely that the combination of the visual features might have led to emergent Gestalt effects that influenced the perception of the overall CAPTCHA figure, which were partially observed in the behavioral data.

We believe that neuroimaging is a promising methodology that will be used for calibrating CAPTCHA configurations in the design of secure software development performed by business units that demands usability and cybersecurity units demanding robustness in the future. These systems have the potential to be neuroadaptive in that the design of a security mechanism may involve self-calibration properties as a function of the user's cognitive workload. Moreover, systematic neuroimaging patterns in CAPTCHA solving tasks may be used for biometric identity verification systems [34–36].

Although the results show systematic data patterns, they are not generalizable at this stage due to multiple factors that need further investigation. In particular, a richer set of visual features is needed to understand their effects on the solution process in isolation and their combination. Nevertheless, that is a challenging task since the combination of visual features is far from being additive. Instead, novel perceptual features emerge from those combinations (namely, emergent features in Gestalt theory of perception), which violates the independence assumptions made by statistical analyses in terms of their calculation of factorial combinations. Accordingly, further studies should address both expanded data sets and novel statistical methods for data analysis. Finally, replicating this study with a full-head fNIRS or fMRI system could help explore the effects of different distortions based on their implications over other related regions in the parietal, occipital, and temporal lobes. Such insights could also help calibrate CAPTCHAs for special populations such as dyslexic users.

## Appendix

## Mean HbO and HbR Responses Observed for Each Main Visual Feature

In each graph, the temporal averages observed for HbO and HbR concentration changes (in $\mu$molar/liter) during the first 30 seconds of each CAPTCHA solving block are shown for all visual feature types. All temporal plots are baseline-corrected and presented on the same scale to aid comparisons, shown in Figures 13–32.

Figures 25–27 comprise lines, Figures 28–30 comprise waves, and Figures 31 and 32 comprise rotation.

## Data Availability

The data that support the findings of this study are available on request from the corresponding author.

## Conflicts of Interest

The authors declare that there are no conflicts of interest regarding the publication of this manuscript.

## Acknowledgments

This research has been funded by Dr. Cengiz Acartürk's (third author) project resources on cybersecurity. The publication costs will also be covered by the researchers' projects.